\newif\ifShowKeys
\ShowKeystrue

\documentclass[11pt]{article}
\pdfoutput=1
 \usepackage[parsep]{collref}

\usepackage[usenames,dvipsnames]{xcolor}
\usepackage[setpagesize=false,pagebackref=false, linktocpage, bookmarksopen=true, colorlinks=true, linkcolor=Maroon,citecolor=Maroon,urlcolor=Maroon]{hyperref}

\usepackage{tocloft}
 
\usepackage{amsmath, amssymb,amsthm}
\usepackage{amsthm}
\numberwithin{equation}{section}

\usepackage{bm,environ,mathrsfs,array,arydshln}
\usepackage{booktabs,float,slashed,hyperref}
\usepackage[mathcal]{euscript}
\usepackage{tensor} 						
\usepackage{mathabx}
\usepackage{simpler-wick}
\usepackage[vcentermath]{youngtab}

\usepackage{aurical}
\usepackage[T1]{fontenc}
\usepackage[nodayofweek]{date time}
\usepackage{graphicx,epsfig,epic}
\usepackage{tikz} 
\usetikzlibrary{arrows,decorations.pathreplacing,decorations.markings,snakes}
\tikzset{middlearrow/.style={decoration={markings, mark= at position 0.5 with {\arrow{#1}} ,
}, postaction={decorate}}}
\tikzset{decoration={snake,amplitude=.4mm,segment length=2mm,
                       post length=0mm,pre length=0mm}}

\usepackage{framed}						
\definecolor{shadecolor}{rgb}{0.9996078, 0.984314, 0.960784}

\usepackage{comment}

\allowdisplaybreaks

\definecolor{myred}{RGB}{233, 33, 45}

\newcommand{\bs}{\begin{shaded}}
\newcommand{\es}{\end{shaded}\noindent}
\def\ba#1\ea{\begin{align}#1\end{align}}		
\newcommand{\be}{\begin{equation}}
\newcommand{\ee}{\end{equation}}
\newcommand{\mc}{\mathcal }

\newcommand{\la}{\label}
\newcommand{\eps}{\varepsilon}

\newcommand{\lp}{\notag \\ & }

\DeclareMathOperator{\tr}{\text{tr}}

\newcommand{\cf}{\textit{cf.} }
\newcommand{\ie}{\textit{i.e.} }
\newcommand{\eg}{\textit{e.g.}}

\newcommand{\N}{\mathcal N}

\makeatletter
\DeclareFontFamily{OMX}{MnSymbolE}{}
\DeclareSymbolFont{MnLargeSymbols}{OMX}{MnSymbolE}{m}{n}
\SetSymbolFont{MnLargeSymbols}{bold}{OMX}{MnSymbolE}{b}{n}
\DeclareFontShape{OMX}{MnSymbolE}{m}{n}{
<-6>  MnSymbolE5
   <6-7>  MnSymbolE6
   <7-8>  MnSymbolE7
   <8-9>  MnSymbolE8
   <9-10> MnSymbolE9
  <10-12> MnSymbolE10
  <12->   MnSymbolE12
}{}
\DeclareFontShape{OMX}{MnSymbolE}{b}{n}{
<-6>  MnSymbolE-Bold5
   <6-7>  MnSymbolE-Bold6
   <7-8>  MnSymbolE-Bold7
   <8-9>  MnSymbolE-Bold8
   <9-10> MnSymbolE-Bold9
  <10-12> MnSymbolE-Bold10
  <12->   MnSymbolE-Bold12
}{}

\let\llangle\@undefined
\let\rrangle\@undefined
\DeclareMathDelimiter{\llangle}{\mathopen}%
 {MnLargeSymbols}{'164}{MnLargeSymbols}{'164}
\DeclareMathDelimiter{\rrangle}{\mathclose}%
 {MnLargeSymbols}{'171}{MnLargeSymbols}{'171}
\makeatother

\catcode`,\active

\catcode`\,12

\def\XXint#1#2#3{{\setbox0=\hbox{$#1{#2#3}{\int}$}
     \vcenter{\hbox{$#2#3$}}\kern-.5\wd0}}

\newcommand{\sql}{\sqrt\l}
\renewcommand{\l}{\lambda}

\newcommand{\rf}[1]{(\ref{#1})}

\newcommand{\OO}{\mathsf{O}}

\newcommand{\hl}{\widehat{\l}}

\newcommand{\gs}{g_{\rm s}}

\def \ov {\over}
\def \ci {\cite}

\DeclareMathOperator{\Ai}{Ai}

\def \adsc {AdS$_4 \times  S^7/Z_k$ }
\def \CP{{\rm CP}}
\def \ha {\tfrac{1}{2}}
\def \foot {\footnote}
\def \T {{\rm T}}  
\def \LL {{\rm L}} 
\def \iffa {\iffalse}
\def \ff {{\rm f}}  \def \ha {\frac{1}{2}}
  \def \llp {\ell_P}
\def \adsz  {AdS$_{4}\times S^{7}/\mathbb{Z}_{k}$ }
\def \adsc  {AdS$_{4}\times {\rm CP}^3$ }
\def \adscd  {AdS$_{4}\times {\rm CP}^3$}
\def \CP {{\rm CP}}
\def \vol {{\rm vol}}
\def \LA  {\Lambda} 

 \def \AA   {{\cal A}}
\def \HH  {{\cal H}}
\def \FF {{\cal F}}
\def \te {\textstyle} \def \ZZ {{\mathbb Z}}
\def \l {\lambda}  \def \ov {\over}

\def \no {\nonumber}
  \def \ta {\tau}
\def \ads {{\rm AdS_4}}
\def \ZZ {{\mathbb Z}}\def \vp {\varphi}
\def \RR  {{\rm R}}\def \td {\tilde}\def \LLL {L_{11}}\def \TT {{\rm T}}

      \def \rs  {{\rm s}}  \def \rt   {{\rm t}}  \def \ru  {{\rm u}}
\def \cL {{\cal L}} 
\def \dd {{\rm d}}
  \def \mm {{\rm  m}}    
\def \half {{\te \ha}}

\def \ed {
\bibliography{BT-Biblio}
\bibliographystyle{JHEP-v2.9}
\end{document}
}

\begin{document}

\begin{titlepage}
\begin{tabbing}
\date{\currenttime}
\hspace*{11.5cm} \=  \kill 
\>  Imperial-TP-AT-2023-03 \\
\> 
\end{tabbing}
\hspace*{1cm}  
\vspace*{15mm}
\begin{center}
{\large\sc   Comments on   ABJM free energy on $S^{3}$ at large $N$\\
\vskip 7pt
 and  perturbative expansions  in M-theory and string theory   }\vskip 9pt
\vspace*{10mm}
{\large M. Beccaria${}^{\,a}$ 
and        A.A. Tseytlin$^{\,b, }$\footnote{Also on leave from the  Institute for Theoretical and Mathematical Physics (ITMP) of MSU   and Lebedev Institute.}} 
\vspace*{4mm}

${}^a$ Universit\`a del Salento, Dipartimento di Matematica e Fisica \textit{Ennio De Giorgi},\\ 
		and I.N.F.N. - sezione di Lecce, Via Arnesano, I-73100 Lecce, Italy
			\vskip 0.1cm
${}^b$ Blackett Laboratory, Imperial College London,  SW7 2AZ, U.K.
			\vskip 0.3cm
\vskip 0.2cm {\small E-mail: \texttt{matteo.beccaria@le.infn.it},\  
\  \texttt{tseytlin@imperial.ac.uk}}
\vspace*{0.8cm}
\end{center}
\begin{abstract}  
\noindent
 We compare
  the  large $N$ expansion  of the localization result for  the  free energy $F$ in the 
 3d $\N=6$   superconformal   $U(N)_k \times U(N)_{-k}$   Chern-Simons-matter
  theory  to its AdS/CFT  counterpart, \ie  to  the perturbative 
  expansion of  M-theory partition function   on \adsz 
   and  to  the weak string coupling expansion of   type IIA  effective action 
  on \adscd. 
  We show  that the  general form of  the perturbative expansions of $F$ 
   on the two sides of  the AdS/CFT  duality  is indeed the same.  Moreover, 
 the    transcendentality  properties of the coefficients in the large $N$,  large $k$   expansion of $F$ match  those in the  corresponding  M-theory or string theory expansions. 
 To  shed light on  the  structure  of the  1-loop  M-theory     partition function on  \adsz 
 we  use the  expression for the 1-loop 4-graviton scattering amplitude in the  11d  supergravity.
 We also use the known information  about the transcendental coefficients of the leading curvature  invariants in the 
 low-energy effective action of type II  string theory.   Matching of the remaining  rational  factors in the coefficients 
  requires a precise information about currently unknown   RR field strength  terms in the corresponding superinvariants. 
\end{abstract}

\vskip 0.5cm	{ }		
\end{titlepage}

\tableofcontents
\vspace{1cm}

\parskip 0.1cm

\def \z  {\zeta} 

\setcounter{footnote}{0}
\section{Introduction}

Localization 
 \cite{Pestun:2007rz,Pestun:2016zxk}  provides a remarkable  source of information about  supersymmetric 
 gauge theories   beyond  the standard weak-coupling perturbation theory.  In the context of  AdS/CFT  duality
 \ci{Aharony:1999ti} this information
 may be used  to learn about  the structure of string theory or M-theory corrections  to  
  the tree level or supergravity order.  
 
 Here we shall focus on the 3d $\N=6$   supersymmetric  $U(N)_k \times U(N)_{-k}$   Chern-Simons-matter 
  theory \ci{Aharony:2008ug}  in which the free 
 energy $F(N,k)$  on $S^3$    was computed by localization in  \ci{Kapustin:2009kz,Marino:2009jd,Drukker:2010nc,Drukker:2011zy} (see \ci{Marino:2016new} for a review and further references). For large $N$ and fixed $k$ this theory is dual to   M-theory  on \adsz  while for  
  large $N$ and large $k$ with fixed $\l= {N\ov k} $ 
  it  is dual  
  to  the 10d type IIA string theory on \adsc  background.\foot{This superconformal theory represents  $N$ M2-branes probing a $\mathbb{C}^{4}/\mathbb{Z}_{k}$ singularity. 
The orbifold acts as $z_{i}\to e^{\frac{2\pi i}{k}}z_{i}$ where $z_{i}$, $i=1,2,3,4$ are four complex
coordinates transverse to the M2-branes.}

  Our aim will be to  compare the large $N$ expansion  of $F$ to its AdS/CFT  counterpart, \ie  to perturbative 
  expansion of the M-theory partition function   on \adsz or weak string coupling expansion of  string theory effective action 
  on \adscd.
   Related   work  appeared in \ci{Bhattacharyya:2012ye,Liu:2016dau}
  and  \cite{Agmon:2017xes,Chester:2018aca,Binder:2018yvd,Binder:2019mpb,Alday:2021ymb,Alday:2022rly}
  and  also in 
  \ci{Bobev:2020egg,Bobev:2021oku,Bobev:2022eus,Bobev:2022jte,Bobev:2022wem}.  
 In the  M-theory  the expansion parameter is the inverse of the  effective  dimensionless M2-brane  tension 
 \ci{Aharony:2008ug}\foot{Here
 we ignore possible ``quantum''  shifts of the parameters   (see below).} 
 \be \la{1} 
 {\rm T_2}  =\frac{1}{(2\pi)^{2}} \frac{L_{11}^3}{\,\ell_{P}^{3}} \ , \ \ \ \  \ \ \ \ \ 
 {L_{11}\ov \llp} = \big(2^{5}\pi^{2} N k \big)^{1/6} \ , \ee
 while   the type IIA string   coupling and effective string  tension    are 
 \ba
\la{2}
&\gs = \sqrt\pi\, \big({2 \ov k} \big)^{5/4}  {N}^{1/4}  =  \frac{\sqrt\pi\, (2\l)^{5/4}}{N} \ , \ \ \ \ \  \ \ \  \ \ \ \  \l= { N \ov k } 
\ , \\
&T = {1\ov 8\pi}  \frac{L^2}{\alpha'} = \gs^{2/3} \frac{\LLL^{2}}{8\pi \alpha'}
= \frac{\sqrt{\l}}{\sqrt 2}\ , \ \ \qquad \qquad 
\ \     \frac{\gs^2 }{8\pi\, T }  ={\l^2 \ov N^2} =  {1\ov k^{2}} \ ,  \la{3}
\ea
where $L_{11}$  and $L$ are curvature scales in the 11d and 10d metrics.

 We will  show  that the  general structure of  perturbative expansions of $F$ 
   on the two sides of the AdS/CFT  duality  is indeed the same. 
    Moreover, 
 the    transcendentality  properties of the coefficients in the large $N$,  large $k$   expansion of the localization 
 expression  for $F$ match  those in the  corresponding  M-theory or string theory expansions. 
   In particular,  we will focus on the $N$-independent
  $A(k)$    part of $F$  and show that   the leading   $\zeta(3) k^2$  term in its large $k$ expansion 
  corresponds to the $\zeta(3)$   term  in the 1-loop   11d graviton  amplitude  on     $M^{10} \times S^1$  \ci{Green:1997as,Russo:1997mk}   or  the  tree-level  $\zeta(3) R^4 $ term in the 10d
   string theory effective action.\foot{In addition to the M-theory and weakly coupled string theory  limits one may 
   consider a  limit of large $N$ with fixed $N/k^5$  
    that corresponds to the  type IIA string at finite string coupling  and thus   interpolates between 
     M-theory at strong coupling and  perturbative  string theory at weak coupling.
        Ref. \ci{Binder:2019mpb} 
      used that  limit to compute $R^4$ terms  at finite coupling.} 
  Also, we  will   find  that the $\pi$-dependent factors  in the coefficients of  subleading $1\ov k^n$ 
  terms   match those in the coefficients  of the corresponding curvature invariants   in the M-theory or string theory effective actions.\foot{This is similar  to what  was observed
      \ci{Beccaria:2023kbl} in 
  the discussion of  the leading strong-coupling  terms  in  the localization result for  free energy in  the orbifold $\N=2$  gauge theory     at each order in  the $1/N^2$  expansion. These terms 
take the form of a series  in ${\lambda^{3/2}\ov N^2}\sim {\gs^2\ov T}$  and    can be matched (up to  rational coefficients)  with the  contributions coming from the  $D^n R^m$ terms (of lowest order in $\alpha'$ at each  order in $\gs^2$)  in type IIB string effective action.}
To  match the  remaining   rational factors in the coefficients requires  precise 
knowledge of the structure of  the corresponding superinvariants (RR flux terms in them) and remains an open problem.

The  order $N^0$   term    in $F$  should   correspond  to the  order $(\rm T_2)^0$ or 1-loop correction 
in M-theory. 
A similar $k$-dependent  factor in the 1/2 BPS  Wilson  loop  expectation value   in the ABJM theory  was recently    reproduced  \ci{Giombi:2023vzu} as the 
  1-loop quantum M2-brane correction.\foot{It was observed in \ci{Beccaria:2020ykg} that  this    $( \sin {2\pi \ov k} )^{-1} $ prefactor (where  ${1\ov k^2} = { \gs^2 \ov  8 \pi T }$ as in \rf{3})  in the  Wilson loop expectation value 
    effectively sums up 
the leading large $T$  contributions at each order in $\gs^2$. 
In \ci{Giombi:2023vzu}  this prefactor was derived as a 1-loop correction in the  M2-brane world-volume  theory
and thus it was concluded  that this 
 1-loop M2  brane correction effectively sums up all 
  large tension  terms at all orders  in  the weak  string coupling expansion in  the dual  type IIA theory.}  
 The $A(k)$ term in the free energy should   represent the contribution 
of  quantum  M2-brane  states  propagating in the loop. In addition to point-like M2 branes  one may 
 need to  include also contributions of BPS M2  branes wrapping 2-cycles  in $\CP^3$ part of  $S^7/\ZZ_k$. 
 As we shall discuss below, the  structure of the $A(k)$   function 
 suggests  a  close analogy of the present case with the  Calabi-Yau compactification one  in \cite{Gopakumar:1998ii,Gopakumar:1998jq,Dedushenko:2014nya}.\foot{
It would be interesting also to try to  do a similar   matching in the case of the topological indices 
(or special partition functions on $S^2 \times S^1$) for which  localization results 
were discussed in \ci{Bobev:2022eus,Bobev:2022jte,Bobev:2022wem}.}

\iffa Here we address  the  same question in the case of ABJM theory  for free energy on $S^3$. 
Expanded at large $N$  it takes the form $F=  c_1 N^{3/2} + c_2 N^{1/2} + ....$ 
and while  $c_1$  was  matched  to   coefficient of $R$ term, the coefficient $c_2$ was not previously matched onto coefficient of $R^4$ term or determined on string side (cf. \ci{Bobev:2020egg,Bobev:2022eus}). 
 This is one open question. 
 For M2  which is point-like  in $CP^3$  this will be  like 1-loop  supergraviton partition function 
 in 11d supergravity  on \adsc  (that  indeed captures $\log N$ term in $F$ \ci{Bhattacharyya:2012ye} 
 but does not give correct result for the  rest of the terms \ci{Liu:2016dau}).
 We may also   consider  4-graviton (or D0 brane) 1-loop amplitude in  flat  11d
 space $M^{10} \times S^1$  \ci{Green:1997as,Russo:1997mk}, extract 
 higher-derivative $D^n R^4$   corrections  from it and then evaluate them on \adsc  background. 
 That seems to fail to precisely match the $A(k)$ term.
 \fi

This  paper is organized as follows. 
In section 2  we review the structure of the large $N$  perturbative part  of  the 
 free energy  as found from localization 
in ABJM theory on $S^3$. 
In section 3 we compare its large $N$, fixed $k$ expansion to  the perturbative expansion of 
the partition function or effective 
action in M-theory. 
In section 3 we  discuss the large $N$, fixed $\l$ expansion of $F$ and show its  correspondence with 
 the perturbative expansion in type IIA string theory on \adscd. 

Some basic relations and notation are summarized in Appendix A. 
In Appendix B we  recall the matching of the leading  large $N$ term in $F$ with the 11d supergravity 
action evaluated on the \adsz   background.  In appendix C we present the \adscd\  values 
of the $R^4$ invariants that appear  in the  tree level  and  one loop term in the type IIA string effective action. 
Appendix D  contains a brief review of the structure of non-perturbative terms in the ABJM   free energy.


\section{Free energy of ABJM  model  in the  large $N$ expansion}

Our starting point will be the localization result for the free   energy of the ABJM theory on $S^3$   expanded  at large $N$. 
We will consider both  fixed $k$ and  large $k$   perturbative expansions  ignoring non-perturbative corrections.

The partition function of the ABJM theory on $S^3$   was  first expressed  
 in terms of a localization matrix model in \cite{Kapustin:2009kz}. 
It was later mapped to a lens space matrix model and solved in  
the planar limit in \cite{Marino:2009jd}. Higher genus $1/N$ corrections were computed
in  \cite{Drukker:2010nc,Drukker:2011zy} by integrating the holomorphic anomaly equation. Neglecting non-perturbative corrections 
(reviewed  in \cite{Hatsuda:2015gca,Marino:2016new})  the resummed partition function was determined in \cite{Fuji:2011km}.
The same result was later rederived by Fermi gas methods in \cite{Marino:2011eh} 
and tested numerically in \cite{Hanada:2012si} at finite $N, k$. 
The resulting perturbative partition function reads 
\ba
\la{2.3}
& Z(N,k) \equiv  e^{- F(N,k)} = (\half { \pi^2 k })^{1/3}\ e^{A(k)}\ \Ai(z),\qquad \qquad   z =  (\half { \pi^2 k })^{1/3}\ \Big(N-\frac{k}{24}-\frac{1}{3k}\Big)\ .  
\ea
The presence of the function $A(k)$  was  first detected in  
\cite{Hanada:2012si}
and  incorporated into the
 Fermi gas formalism of  \cite{Marino:2011eh}
 that  provided its  small $k$ expansion. 
The large $k$ expansion of $A(k)$ was identified  in  \cite{Hanada:2012si}  with a topological string "constant map" contribution  \cite{Bershadsky:1993cx,Faber:1998gsw}.\foot{The lens space Chern-Simons matrix model  partition function 
 can be interpreted as a partition  function of  a large $N$ dual 
 of  a  topological string theory on a certain class of local Calabi-Yau geometries \cite{Aganagic:2002wv}.
This is a generalization of the Gopakumar-Vafa duality \cite{Gopakumar:1998ki}.
}
Ref. \cite{Hanada:2012si}  proposed a resummed integral representation for $A(k)$ (later improved in \cite{Hatsuda:2014vsa}) 
valid at both small and large $k$
\foot{The specific values of $A(k)$ 
 at  integer $k$ are given in Eq.~(3.14) of  \cite{Hatsuda:2014vsa}. 
In particular,  $A(1) = -\frac{\zeta(3)}{8\pi^{2}}+\frac{1}{4}\log 2$.
} 
\ba
\la{255}
A(k) &=- \frac{\zeta(3)}{8\pi^{2}}\Big( k^2  - {16\ov k}\Big)+\frac{k^{2}}{\pi^{2}}\int_{0}^{\infty}dx\frac{x}{e^{kx}-1}\log(1-e^{-2x})\ .
\ea
The 
expansions of (\ref{255}) in the two regimes may be determined as asymptotic series. For  $k\ll 1  $ 
\ba
\la{233}
A(k) &\stackrel{k\ll1 }{=}  \ \ \frac{2\, \zeta(3)}{\pi^{2}}{1\ov k}+\sum_{n=1}^{\infty}(-1)^{n}\frac{\pi^{2n-2}}{(2n)!}\,B_{2n}B_{2n-2}\, k^{2n-1}\ ,
\ea
where $B_{2n}$ are Bernoulli numbers. At large $k$  one finds  
\ba
\la{2.5}
A(k) &\stackrel{k\gg 1}{=} -\frac{\zeta(3)}{8\pi^{2}}\,k^{2} 
+\frac{1}{6}\log\frac{4\pi}{k} + 2\zeta'(-1)  + \bar A(k) \ , \qquad \qquad \bar A(k) = \sum_{h=2}^{\infty} \frac{q_h}{k^{2h-2}}\ , 
\ea
where $q_n$ are   rational numbers    expressed again in terms of  the 
products of two Bernoulli numbers  or even-argument zeta-function values
$\zeta(2n) = (-1)^{n+1} {( 2 \pi)^{2n}  \over 2  ( 2 n)!}B_{2n}$ as 
\ba
\la{2.6}
q_{h} =& \frac{(2\pi)^{2h-2} (-1)^{h+1}4^{h-1}}{h(2h-2)(2h-2)!}\, B_{2h}\, B_{2h-2}\  . 
\ea
The expansion (\ref{2.5}) reproduces (see below)  the dominant terms  in $\l=\frac{N}{k}\gg 1$ in the 
$1/N$ expansion of \rf{2.3}. 
For this reason the resummation proposal (\ref{255}) is usually considered to be correct. \footnote{
Numerical tests of (\ref{255}) for intermediate values of $k$ were also presented in \cite{Hanada:2012si}.}

Below we shall use \rf{2.3},\rf{255} as a starting point ignoring non-perturbative corrections
(for some comments on them see Appendix \ref{D}).

From the exact expression of $F(N,k)$ we can work out  its large $N$ expansion at fixed $k$
\ba
\la{2.7}
F =& \frac{1}{3} \sqrt{2}  \pi  k^{1/2}  N^{3/2}-\frac{\pi}{24\sqrt 2}\big( k^2 +8 \big)\,k^{-1/2}\,N^{1/2}  
 +\frac{1}{4}  
\log  {32  N\ov k}  - A(k) \lp+\frac{\pi (k^2+8 )^2 }{2304 \sqrt{2} 
k^{3/2}  {{N^{1/2}}}}
-\frac{k^2+8}{96 k N}+\frac{69120 k^2+   \pi^2 (k^2 + 8 )^3
}{331776 \sqrt{2} 
k^{5/2} \pi  {N}^{3/2}}
-\frac{(k^2 +8)^2}{4608 k^2 N^2} + \cdots \, .
\ea
One can then  assume that $k$ is large and isolate    the leading terms in  $k$    order by order in large $N$
\ba\la{28}
F^{k\gg 1} =& \frac{\sqrt 2 \pi}{3}\, k^{1/2}\, N^{3/2}-\frac{\pi}{24\sqrt 2}k^{3/2}\,N^{1/2} 
+\frac{1}{4}\log N +\frac{\zeta(3)}{8\pi^{2}}k^{2} \lp
+\frac{\pi}{2304\sqrt 2}k^{5/2}N^{-1/2}-\frac{1}{96}kN^{-1}+\frac{\pi}{331776\sqrt 2}k^{7/2}N^{-3/2}+\cdots\ .
\ea
Here the $\zeta(3)$ term came  from  
  the first term in (\ref{2.5}). 
As follows from \rf{2.3}, 
for  large $k$ at each order in $1/N$ the relevant combination   should 
 be $N-{1\ov 24} k $  and indeed   one finds  that \rf{28}  may be rewritten as 
\ba
\la{2.9}
F^{k\gg 1} &
= \frac{\pi}{3} \sqrt{2k}  \Big(N-\frac{k}{24}\Big)^{3/2}  
+\frac{1}{4} \log\Big(N-\frac{k}{24}\Big)   +\frac{\zeta(3)}{8\pi^{2}}k^{2}  + \cdots\, .
\ea
In the 't Hooft  expansion, \ie  the expansion in $1/N$  
 with fixed $\l= {N \ov k} $,  the resulting large $N$ expression  of $F$  may be written as  
\be
\la{2.10}
F = -\log Z = -\sum_{h=0}^{\infty} 
 (-1)^{h-1} f_{h}(\l)\Big( \frac{2\pi \l}{N}\Big)^{2h-2}  + {1\ov 6} \log {N\ov \l}\ . 
\ee
Since we isolated in \rf{2.10}  the  $1/N$   factors  in the combination $ {\l\ov N} = {1\ov k}$, 
it follows from (\ref{2.3}) that  the functions $f_{h}(\l) $  should naturally  depend  on the shifted coupling
\be
\la{2.11}
\hl \equiv  \l-\frac{1}{24} = {1\ov k} ( N - \frac{1}{24}  k) \ .
\ee
Explicitly, one  finds  the following expressions for $f_n$  
 \cite{Bobev:2022eus} ($\mathsf{A}$    is  Glaisher constant)
\ba
 f_{0} =&\te  \frac{4\sqrt 2\,\pi^{3}}{3}\hl^{3/2}+\frac{1}{2}\zeta(3)\ , \quad \qquad  
f_{1} =\te  \frac{\pi}{3\sqrt 2}\hl^{1/2}-\frac{1}{4}\log\hl 
+\frac{1}{6}-\frac{11}{12}\log 2+\frac{1}{6}\log\pi-2\log\mathsf{A}\ , \no  \\
f_{2} =&\te  -\frac{1}{360}+\frac{1}{144\pi\sqrt 2}\hl^{-1/2}-\frac{1}{48\pi^{2}}\hl^{-1}+\frac{5}{96\pi^{3}\sqrt 2}\hl^{-3/2}, \no \\
f_{3} =&\te  -\frac{1}{22680}-\frac{1}{10368 \sqrt{2} \pi ^3}\hl^{-3/2}
+\frac{1}{1152 \pi ^4 }\hl^{-2}-\frac{5}{768 \sqrt{2} \pi ^5}\hl^{-5/2}+\frac{5}{512 \pi ^6}\hl^{-3},  \la{2.100}  \\
f_{4} =&\te  -\frac{1}{340200}+\frac{1}{331776 \sqrt{2} \pi 
^5}\hl^{-5/2}-\frac{1}{20736 \pi ^6 }\hl^{-3}+\frac{25 }{36864 
\sqrt{2} \pi ^7}\hl^{-7/2} 
-\frac{5}{2048 \pi ^8}\hl^{-4}+\frac{1105 
}{147456 \sqrt{2} \pi ^9}\hl^{-9/2},\no  \\
f_{5} =&\te  -\frac{1}{2494800}-\frac{1}{7962624 \sqrt{2} \pi 
^7}\hl^{-7/2}+\frac{1}{331776 \pi ^8 }\hl^{-4}-\frac{175 
}{2654208 \sqrt{2} \pi ^9}\hl^{-9/2}\lp\te 
\qquad \qquad\ \  +\frac{5}{12288 \pi 
^{10}}\hl^{-5}-\frac{1105 }{393216 \sqrt{2} \pi 
^{11}}\hl^{-11/2}+\frac{565}{131072 \pi ^{12}}\hl^{-6}\ ,\ \  ...\no 
\ea
A remarkable feature of these expressions  for $f_h(\hl)$ is that  they are  given  by  finite sums of terms. 
For  $h\geq 2$  we get 
\be\la{2.18}
f_{h} (\hl)  = p_{h}+\sum_{s=1}^{h+1} \frac{ p_{h,s} }{\big( 2\pi\, \sqrt{\hl}\ \big)^{s + 2h+4}}\ ,\qquad \ \ \  h \geq 2 \ , 
\ee
where all $p_h$ and $p_{h,s}$  are 
{\em rational} coefficients.

As a  further refinement, we may consider the $\l\gg 1$   expansion   and 
isolate the leading powers of   $\l$ at each  order in  $1/N$ 
 in  (\ref{2.10}). These special terms read (omitting $\log k = \log {N\ov \l}$ term in \rf{2.10}) 
\ba
\la{2188}
\widetilde{F} &\equiv F^{\l\gg 1} =  N^{2}\,\Big(\frac{\pi\sqrt{2}}{3}\l^{-1/2}-\frac{\pi\sqrt{2}}{48}\l^{-3/2}
+\frac{1}{8\pi^{2}}\zeta(3)\l^{-2}+...\Big)-\frac{\pi}{3\sqrt{2}}\l^{1/2} \ +\  \bar F  \ ,  \\
\bar F &= - \bar A(k)   =  -\sum_{h=2}^{\infty} 
 (-1)^{h-1} p_{h} \Big( \frac{2\pi }{k}\Big)^{2h-2} =  -\sum_{h=2}^{\infty} 
 (-1)^{h-1} p_{h} \Big( \frac{2\pi \l}{N}\Big)^{2h-2}    \no \\
& \qquad \qquad\  = -\frac{\pi^{2}}{90} \frac{\l^{2}}{N^{2}}+\frac{2\pi^{4}}{2835}\,\frac{\l^{4}}{N^{4}}
-\frac{8\pi^{6}}{42525} \frac{\l^{6}}{N^{6}}+\frac{16\pi^{8}}{155925} \frac{\l^{8}}{N^{8}}+\cdots\ . \la{2.19}
\ea
Here   we kept  few     subleading large $\l$ terms  only in the first $N^2$ term. 
Comparing to \rf{2.7},\rf{2.5}   we conclude that 
the coefficients  $p_h$ are related to $q_h$ in $\bar A$ in  \rf{2.5} as (cf. \rf{2.6})
\be 
p_h = { (-1)^{h-1} \ov  (2 \pi)^{2h-2} 
} \, q_h    =  \frac{4^{h-1}B_{2h}\, B_{2h-2}  }{  h(2h-2)(2h-2)!  }  \ . \la{215}\ee

The two expansions
we have discussed  (large $N$ at fixed $k$ and  large $N$, large $k$  with fixed $\l= {N \ov k} $) 
should correspond to the M-theory     and   type IIA  string theory  expansions.  
We shall discuss this connection in the next sections. 

\section{
M-theory  perturbative  expansion}





The  large $N$, fixed $k$  expansion of  the ABJM   theory should be 
 dual to  the    perturbative  expansion of  M-theory on \adsz 
 in which   the  curvature scale $L_{11}$  is  small compared to the  11d Planck length $\ell_{P}$   so that the    effective 
 dimensionless M2-brane   tension $\T_{2}$  is large  (see Appendix \ref{apA}  for our notation)
\be\la{3.1}
\T_{2} \equiv L_{11}^{3}  T_2=  \frac{L_{11}^{3}}{(2\pi)^{2}\ell_{P}^{3}} = \frac{\LL^{3}}{4\pi^{2}}\ , \qquad\qquad  \LL = \frac{L_{11}}{\ell_{P}}\gg 1 \ ,
\ee
while the parameter $k$ of the 11d  background (related to the radius of the 11d circle) is fixed. 
Indeed, since  according to  \rf{A.9}
\be \la{311} 
 \LL^6=   32 \pi^2  Nk  \ ,  \ee 
 this limit  is equivalent to the large $N$,  fixed $k$ expansion. 
 
 Thus the M-theory perturbative expansion should    be in  inverse powers of $\T_2$  or in powers of $\LL^{-3}$. 
 Expressed in terms of  $\LL$   and $k$  the large $N$ expansion of  the free energy $F$ in (\ref{2.7}) 
  is indeed  
\ba
\la{3.2}
F =& c_{0}\frac{1}{k}\,\LL^{9}+c_{1} \frac{8 + k^2}{ k}\LL^{3} +   \frac{1}{2}\log {{\LL^{3}}\ov \pi k } 
  - A(k)    + c_{2}\frac{(8+k^{2})^{2}}{k}\LL^{-3} +  c_{3}\frac{8+k^{2}}{k}\LL^{-6}  
    +\OO (\LL^{-9}), \\
& \ \  c_{0} = \frac{1}{384\pi^{2}}, \qquad c_{1} = -\frac{1}{192}\ , \qquad c_{2} = \frac{\pi^{2}}{576}, \qquad c_{3} = -\frac{\pi^{2}}{3}
 \ , \ \   ... \la{3.4}
\ea
As was suggested in  \ci{Bergman:2009zh}, the presence of the  $\int R^4\, C_{3}$ term  in the 11d effective action \cite{Duff:1995wd} implies  the following  shift  of the M2-brane charge $N$  
 \be   \la{32} 
 N \to    N - {1\ov 24} (k - k^{-1}) \ .  \ee
This   leads to  the following  redefinition of $\LL$ in \rf{31}     \ci{Drukker:2011zy} 
\be \la{31} 
 \LL^{6}=  32 \pi^2  \big[Nk  - {1\ov 24} (k^2 - 1) \big] \ .  \ee 
Expressing  the localization result for $F(N,k) $   in \rf{2.7} in terms of  this redefined  parameter $\LL$ and $k$ we find a 
remarkable simplification of the $k$-dependent  coefficients of   the  $\LL^{3}$   powers   
\ba
\la{322}
F = &c_0  \frac{1}{k}\,\LL^{9}  + c_1'  \frac{1}{k}\LL^{3} +    \frac{1}{2}\log {{\LL^{3}}\ov \pi k }  
  - A(k)    +c_{2}' \frac{1}{k}\LL^{-3}+      c'_{3} \LL^{-6}  +   \OO (\LL^{-9}), \\
  & c_{1}' = -\frac{3}{64}, \qquad\ \ \  c_{2}' = \frac{9\pi^{2}}{64}\ , \qquad \ \    c_3'= - 3 \pi^2 \ . \la{39}
   \ea 
  Thus 
  the $k$-dependence of  the $\LL^9, \LL^3$ and $\LL^{-3}$ terms becomes simply $1\ov k$
  (though this  does not apply to  $\LL^{-6}$  and   higher order terms in the expansion). 
  
It  is   natural to   expect  that  the 
 terms  in the free energy  that scale as $1\ov k$    may  originate  from local  terms in  the  M-theory 
partition function or the effective action evaluated on the \adsz  background. 
Indeed,  as  this  background  is  homogeneous 
(and its curvature does not depend on $k$ explicitly, apart from the  dependence  via $L_{11}$ or $\LL$) 
  the  integrals of curvature (and 4-form)  invariants will  be proportional to the factor of the 
 radius $a= {1\ov k}$  of the 11d  circle  coming   from the integration volume. 
  Other 
 terms that   do not scale as $1\ov k$   may  come from 
 non-local  contributions to  the  M-theory partition function.


 \subsection{Local     terms   }
 \la{sec:local}
 
The   $\LL^{9}$ term in \rf{322}   comes   effectively  from the 11d supergravity 
action    $S_0=\frac{1}{2\kappa_{11}^{2}}\int  d^{11} x \sqrt {-G}  ( R + \cdots)$  in \rf{A.1} evaluated on the \adsz    background. 
The value  of the coefficient $c_{0}$  is   reproduced   
 after  taking into account  the regularized value of the volume of AdS$_4$  \ci{Drukker:2010nc,Herzog:2010hf}  
(see Appendix \ref{app:EH}).  In particular,  using that 
$R\sim  (L_{11})^{-2}$  and (extracting the  overall $(\ha L_{11})^{4} (L_{11})^7$
scale factor of  the 11d volume, see  \rf{B.2},\rf{B.3}) 
\be \la{38}
\vol \big({\rm AdS_4} \times S^{7}/\mathbb{Z}_{k} \big) ={4\pi^2\ov 3}  \times   {\pi^4\ov 3}\,   {1\ov k}    \ , 
\ee
and  also that   $2\kappa^2_{11} = (2\pi)^8 \llp^9$ (see \rf{A.1})   we conclude 
that the  coefficient of the   supergravity  term should 
scale as ${1\ov \pi^2}   {1\ov k} \LL^9$  matching  the $\pi^{-2} $-dependence of  $c_0$ in   \rf{3.4}.

 Similarly,  the  ${1\ov k} \LL^{3}$  term in \rf{322} should come 
 from  the local  1-loop  $R^4+\cdots$    term   in the 11d effective action 
 \ci{Green:1997di,Green:1997as,Russo:1997mk,Tseytlin:2000sf}\foot{Note that 
  here our  $ \ell_{P}$    (see Appendix \ref{apA} for  notation)
 is related to $\ell_{11}$  used  in \ci{Russo:1997mk,Tseytlin:2000sf}  as $\ell^3_{11} = 2 \pi \llp^3$ 
 so that  the  values of $\kappa_{11}$  and M2-brane tension $T_2$ are   the same as in these papers. 
 }
 \be \la{3.8}
 S_1=  \,  b_1  T_2   \int  d^{11} x \sqrt {-G}   \big( R^{4} + ...\big) \ , \ \ \ \ \  \ \ \ T_2 = {1\ov (2\pi)^2 \ell_{P}^{3}} \ , \qquad 
  b_1= { 1\ov  9\cdot  2^{13} \cdot  (2 \pi)^4} \ .   \ee
 Here we isolated the factor of  the M2 brane tension $T_2$.
 This  term may be viewed as 
   the 1-loop 11d   supergravity  contribution $\Lambda^3 R^4+\cdots$  that   scales 
 as $\kappa_{11}^0$ but is   cubically divergent   \ci{Fradkin:1982kf}    leading to a finite term in   \rf{3.8}    after  assuming 
 the  M-theory UV cutoff $\L \sim \llp^{-1}$.\foot{This $R^4$  term should  be a superpartner of the $R^4 C_3$ term.
 The fact that accounting for the shift \rf{31} 
 removes the  "non-local" $k \LL^3$ term in \rf{3.2}  may  be  viewed as a   consequence  of supersymmetry.
 Note also that in general
   higher loop  supergravity contributions should   scale as $(\kappa^2_{11})^{L-1} \sim  (T_2)^{-3(L-1)} $  but  
 in local terms  extra  factors of   the M-theory UV cutoff $\L \sim \llp^{-1}$   may introduce  extra  positive 
 powers  of $T_2$,  see  \ci{Russo:1997mk, Green:1999pu} and Eq. \rf{398}  below.
 }
 Thus  this local   1-loop $R^4$   term  is the one that  corresponds to  the $k^{-1/2} N^{1/2} $ term in $F$ in \rf{2.7}.

Let us  recall    that  similar  terms $\sim N^{3/2}$  and $\sim N^{1/2}$  
  appear   in the finite temperature  free energy  of the world-volume theory of multiple   M2 branes  
and   have similar origins  in the  $R$   \cite{Klebanov:1996un}  and $R^4$  \cite{Gubser:1998nz}   
  terms in the M-theory  effective action.


To reproduce the  value of the 
 coefficient $c'_1$  in  \rf{322},\rf{39}  one needs the  information  about    the precise  structure of the 4-form dependent terms in the 
  $R^4$  superinvariant  
which is  not yet known   (cf. \ci{Tseytlin:2000sf}).\footnote{While  matching the overall coefficient  $c_1'$   is thus an open problem, 
in \cite{Bobev:2020egg,Bobev:2021oku} 
the dependence of the  coefficient  of the  similar $N^{1/2} \sim \LL^3$   term 
  on extra  geometric  parameters  (like squashing of the $S^3$) in the localization result  for the free energy  $F$ 
was reproduced from  the  effective 4d  effective action with  the supersymmetric   $R^2$ terms 
that should  originate  from  the 11d   $R^{4}$ superinvariant   compactification to 4d.}
 Still, it is remarkable that 
 the fact  that   the value of  $c_1'$  
that comes from   \rf{3.8}   on the \adsz   background 
is  rational  as in \rf{39} does follow from the  values of $b_1$  in \rf{3.8}   and  of the volume factor \rf{38}:  all 
   factors of $\pi$  cancel out. 
  
In general,  on dimensional grounds, 
all    local terms   in the M-theory  effective action should  contain particular powers  of 
 the  M2-brane tension, \ie   should    be given by the sum of terms like  \ci{Russo:1997mk}\foot{The special  role of
   the terms \rf{398}   noted in  \ci{Russo:1997mk} is that  upon reduction to 10d 
they    have perturbative dependence on  the string coupling $\gs$.  Note that some of these terms   may be interpreted as 
higher-loop  corrections  in 11d  supergravity  proportional to  
$ (\kappa^2_{11})^{L-1} \LA^{3n} \sim   (T_2)^{3-3L  + n }$  
 ($\LA\sim \llp^{-1}$ 
is the 11d UV cutoff).} 
\be \la{398}
S_p=     T_2 ^{3-2p }    \int  d^{11} x \sqrt {-G}   \Big[  b_p  (D^2)^{3p-3}  R^4  + \td b_p R^{3p+1} +\cdots  \Big] \ , 
\ee
where   dots stand for other possible  terms (depending also  on $F_4$) that  have  the same  mass dimension $6p+2$.
Explicitly, the $S_1$ in  \rf{3.2}  corresponds to the $p=1$ case  of \rf{398},  
 the  $p=2$   case  is 
 $ S_2=  \,   T_2 ^{-1}    \int  d^{11} x \sqrt {-G}   \big( b_2  D^6 R^{4} +  \td b_2  R^7 + \cdots\big)$, etc. 

Evaluated on \adsz  background \rf{398}  will scale   as ${1\ov k} \LL^{9-6p}$  and may, in principle, match some of the  subleading terms in  in the free energy \rf{322}.   
Terms   that do not scale as $1\ov k$  should come from non-local  parts of the  quantum  M-theory  effective action.

\subsection{
Terms   corresponding to the  1-loop  M-theory contribution  } 


The terms  $ \ha \log{ \LL^3\ov \pi k }$ and $ -A(k)$  in  \rf{3.2}  which are  of 
zeroth order in  the effective M2-brane  tension \rf{3.1} 
  should   originate from    the  (UV finite part of)  1-loop  contribution  to the  M-theory  partition function.
The  logarithmic $ {3\ov 2} \log \LL$ term  coming from 
 $\frac{1}{4}\log {32 N \ov k} = \frac{1}{2}\log \frac{\LL^{3}}{  \pi k}$ term   in \rf{2.7}  
was  reproduced  by  a 1-loop computation in 11d supergravity in \cite{Bhattacharyya:2012ye} 
as a  universal contribution of   the zero modes  
of the  11d supergravity fluctuation operators on   AdS$_{4}\times X^{7}$  background (with the dependence on $\ell_P$ 
via  $\LL$  coming from normalization factor  related to $\kappa_{11}$). 
 The  $-\ha \log (\pi k)$ term    should have a  similar  origin 
  (being  also related to the volume factor in the normalization of   the supergravity modes).

The   
 $- A(k)$ term in \rf{3.2}  (see  \rf{2.7},\rf{2.5})     
  should  correspond to  the    $\LL$-independent part of the 
 1-loop  contribution  in M-theory on  AdS$_{4}\times S^{7}/\mathbb{Z}_{k}$.
 
 In general, the  1-loop M-theory   partition function     should     be  the contribution of 
virtual  M2-brane propagating in the loop  but it is not clear how to define  it    precisely. 
 In the case of  a  large amount of supersymmetry  of the background
 one may  conjecture  that   only   special   BPS   states  (\eg corresponding to M2-branes
  wrapped on special 2-cycles of  internal space)
  may be   contributing to the  1-loop partition 
  function, 
  while contributions of non-BPS states  may  cancel due to extended supersymmetry of the background
  (cf. \cite{Gopakumar:1998ii,Gopakumar:1998jq,Dedushenko:2014nya}).

  One  may   start with  the contribution of just   point-like  BPS  states   corresponding to 
      the 11d supergravitons, \ie  approximate   the M-theory 
      1-loop  partition   function by its 11d supergravity counterpart.
      To get an insight about the structure of the latter  
       and to compare it  with    $F$  in \rf{322} 
        we will   be guided   by  the expression for the 
   low-energy expansion of   the 1-loop correction   
  to the 4-graviton amplitude  in  11d supergravity \ci{Russo:1997mk}.
  While there is no a priori reason  why  just the supergravity  correction  should be enough to capture  the full M-theory result, 
  we will show that  it  indeed reproduces the structure of the large $k$ expansion of  the corresponding term in $F$.\foot{Let us note    that ref. 
\cite{Liu:2016dau}  attempted (unsuccessfully)  to reproduce  the  leading  large $k$ terms
in   the localization  expression   for the  function  $A(k)$  in \rf{3.4},\rf{2.5}
 by   the 1-loop  computation in the 11d  supergravity
 on  AdS$_{4}\times S^{7}/\mathbb{Z}_{k}$  
  explicitly accounting for the  contribution of the tower of all  11d supergravity 
  KK modes on $S^{7}/\mathbb{Z}_{k}$. 
 The  computation involved  several subtle points that remain to be sorted out.
In particular, it  is also possible that  
that one needs a special  regularization  (consistent with 11d symmetry) 
 different from the one used in \ci{Liu:2016dau}.}

  Our strategy  will be as follows. We shall   consider  the expression   for 
  the 1-loop 4-graviton  amplitude  in 11d supergravity expanded 
    near flat  space  with 11d circle   of radius $R_{11}$   (found under  a  simplifying assumption that 
    only 10d  components of the 4  polarization tensors  and external momenta  are non-zero)
    following  \ci{Russo:1997mk}.
    We shall then expand  this amplitude in powers of momenta and  
    extract its  dependence on  $R_{11}$  and 11d  UV cutoff $\LA\sim \llp^{-1} $.   
    Finally, we will assume  that it can be used to shed light on the structure of 11d   supergravity 1-loop 
     partition function on a curved background. 
     Specifying to the case of  the 
      \adsz background  we shall  reproduce the  structure of the 
      $\LL^3 , \   \log \LL$ and   $  A(k)$  terms in \rf{322}. 
       Remarkably, we shall 
      find the    terms  with the same   transcendental  
      coefficients $\zeta(3)$ and $\pi^{2h-2}$  that appear in the large $k$ expansion of $A(k)$ in \rf{2.5}.\foot{Let us note that in    
        refs.  \ci{Alday:2021ymb,Alday:2022rly}  the values of 
       the function $A(k)$ and its second and fourth 
derivatives at $k=1$ and $k=2$ where related to  the M-theory  4-graviton  1-loop amplitude and were shown to be
        consistent with  the  coefficient of the $R^4$ term in the effective action.}  
      
     In order to match the remaining rational factors in the coefficients it appears   that 
        one is  to include other  contributions to the M-theory partition function on \adsz background. 
        These are  presumably   of  other (extended) BPS M2-brane states propagating 
      in the loop.  
       By analogy with the case of the Calabi-Yau compactification  \cite{Gopakumar:1998jq}
      we  
       shall then discuss   how one could 
      try to modify 
        the supergravity-based result  
         in order to  reproduce  the  double-Bernoulli structure of the coefficients in $A(k)$  in 
       \rf{2.5},\rf{2.6}.


  The  4-graviton  amplitude may be written as
   (omitting  polarization tensor  and   normalization  factors including 10d volume  and momentum delta-function)   \ci{Green:1982sw,Green:1997as, Russo:1997mk}  
  \ba
\la{313}
\hat  \AA_4(\rs,\rt)= 
\AA_4(\rs,\rt)  +  (\text{symm in } \rs,\rt,\ru) ,   \qquad \qquad 
\AA_4(\rs,\rt) =& \sum_{n=-\infty}^\infty\int_{\LA^{-2}}^{\infty}\frac{d\tau}{\tau^{2}}\ e^{-\frac{\tau\,  n^{2}}{R_{11}^{2}} }\ P(\rs,\rt; \tau), 
\ea
where $\rs,\rt,\ru$ are the  standard kinematic  variables depending on 10d momenta, 
   the sum is over 11d  component of the  virtual momentum  and 
   $\tau$   has dimension of length squared. 
The  function $P$   is given by\footnote{Here we redefined $\tau$ by $\pi$  compared to \ci{Russo:1997mk}
so that it has  direct proper-time interpretation. 
Note that  the factor $\pi$ in front of $M(\rs,\rt; \rho)$    was 
 omitted in going from the first to second line of Eq.~(C.7) in \cite{Green:1982sw}
 and  as a result was missing in  the expression given in \ci{Russo:1997mk}.}
\ba
P(\rs,\rt; \tau) =& \int d^3 \rho \ e^{-\, \tau\,  M(\rs,\rt; \rho)}\ , \qquad \qquad \int d^3 \rho \equiv \int_{0}^{1}d\rho_{3}\int_{0}^{\rho_{3}}d\rho_{2}\int_{0}^{\rho_{2}}d\rho_{1}
 \ ,\la{3143} \\
M(\rs,\rt;\rho) \equiv & \ \rs\rho_{1}\rho_{2}+\rt\rho_{2}\rho_{3}+\ru\rho_{1}\rho_{3}+\rt(\rho_{1}-\rho_{2}), \qquad\ \ \  \rs+\rt+\ru=0.
\label{C.5} 
\ea
Focussing on the first term in the sum in  \rf{313}   and expanding  $e^{- \tau M}$ in powers of  momenta, or, equivalently, in powers of $M$ we get
\ba
& \AA_4(\rs,\rt) =  \sum_{h=0}^{\infty}\AA_{4,h}(\rs,\rt) \ , \la{513}
 \qquad\ \ \  \ \ \ \AA_{4,h} = \sum^\infty_{n=-\infty}\int_{\LA^{-2}}^{\infty}\frac{d\tau}{\tau^{2}}\  e^{-\frac{\tau\,  n^{2}}{R_{11}^{2}}} \ 
\frac{(-1)^{h}}{h!}\,  \tau^{h} \ H_h(\rs,\rt) \ ,\\  & \qquad \ \ \
H_h\equiv   \int d^3 \rho \, M^{h}(\rs,\rt; \rho ) = s^h \bar H_h \big({\rs\ov \rt}\big)\, .\la{315}
\ea
The  $h=0$ term in \rf{513}  may be written  (using Poisson resummation and $H_0 = {1\ov 6}$) as 
 \cite{Green:1997as}
\be\la{316} 
\AA_{4,0} = \frac{2}{3\pi }R_{11}\Lambda^{3}+\frac{\zeta(3)}{\pi^2 R_{11}^{2}}\ .
\ee
Here the  first term comes effectively
   from the $n=0$  contribution  and is thus   the same as in the 
1-loop  contribution in 10d supergravity. The second term comes from  the contribution of   11d  supergravity states  with non-zero 11d momentum (or,  from the 10d string theory  point of view,  from the  contribution of the 
massive D0-brane states  in the loop \ci{Green:1997as}).    
The $h=1$ contribution    vanishes after integrating over $\rho$,  in agreement with 
the absence of 1-loop logarithmic divergences in 11d theory (and also  in the 
 1-loop 4-graviton amplitude in 10d supergravity \cite{Metsaev:1987ju}).

 The remaining $h\ge 2$ terms  are UV finite.  The $n=0$  term in the sum in \rf{315}  with $h \geq 2 $ 
   gives a non-analytic  contribution   ($\sim \rs \log \rs$, \textit{etc.}) 
   to  \rf{513}  
  which is independent of $R_{11}$ (and thus  should be the same as  the 1-loop amplitude in 10d supergravity)
 \be 
 \HH (\rs,\rt)=  \sum_{h=2}^\infty  \int_{0}^{\infty}\frac{d\tau}{\tau^{2}}\  \frac{(-1)^{h}}{h!}\,  \tau^{h}\,   H_h(\rs,\rt) 
=   \int d^3 \rho   \, M (\rs,\rt; \rho ) \log M (\rs,\rt; \rho )   \equiv   \rs \, \bar \HH \big({\rs\ov \rt}\big) 
   \ . \la{815}
\ee
The  contribution of the $n\neq 0 $ terms   
  may be written as 
\ba
\la{C.10}
\sum_{h=2}^{\infty}  \AA'_{4,h}(\rs,\rt) =& 2\sum_{n=1}^{\infty}\int_{0}^{\infty}\frac{d\tau}{\tau^{2}}e^{-\frac{\tau n^{2}}{R_{11}^{2}}}\,\sum_{h=2}^{\infty}\frac{(-1)^{h}}{h!} \,\tau^{h}\, H_{h}(\rs,\rt) \equiv \sum_{h=2}^{\infty}  C_h\,  H_{h} (\rs,\rt)    \ ,  
\ea
where    $C_h$ 
    is given by 
\ba
 & C_h= 2\,\frac{(-1)^{h}}{h!}  \sum_{n=1}^{\infty}\int_{0}^{\infty}\frac{d\tau}{\tau^{2}}e^{-\frac{\tau n^{2}}{R_{11}^{2}}}\,\tau^{h} = 2\frac{ (-1)^{h}}{h!}  \sum_{n=1}^{\infty}(h-2)!\, \Big(\frac{n^{2}}{R_{11}^{2}}\Big)^{1-h}
 =  \dd_h R_{11}^{2(h-1)}, \la{319} \\
&\qquad \qquad \qquad  \dd_{h}=\frac{2(-1)^{h}}{h(h-1)}\zeta(2h-2) = \frac{(2\pi)^{2h-2}\, B_{2h-2}}{h(h-1)(2h-2)!} \ ,  \la{4.2} 
\ea
where $B_{2h-2}$ are Bernoulli numbers. 
Adding to \rf{C.10}   the $h=0$ term  \rf{316}  and    the non-analytic  $\HH $  \rf{815} contribution gives 
\ba
\la{4.1}
\AA_4(\rs,\rt) &={2\ov 3\pi } R_{11} \LA^3     + \rs\, \bar {\HH}\big({\rs\ov \rt }\big)  +  \bar \AA_4(\rs,\rt)\ ,\\
  \bar \AA_4(\rs,\rt)& =  \frac{\zeta(3)}{\pi^2 R_{11}^{2}}  
+\sum_{h=2}^{\infty}d_{h}\, R_{11}^{2(h-1)}\, 
\rs^{h}\bar H_{h}\big({\rs\ov \rt} \big) \ . \la{411}
\ea
Here $\bar{\HH}$   contains log terms while  $\bar H_{h}$ are polynomials   of degree $h$. Note that    
all the  terms in \rf{4.1}   have  the same dimension  (length)$^{-2}$.
In   \rf{4.1} we   separated  the first term that  is the only one that depends on $\llp$ via $\LA$.

Let  us   now interpret \rf{4.1} as providing   an   indication  about 
 the structure  of the M-theory 
 1-loop partition   function   on a curved background. 
 Specializing to  \adsz  we will   have the 11d radius 
 $R_{11} \to  {1\ov k} L_{11}$  and, just on dimensional grounds,   the momentum variables  $\rs,\rt$ scaling as $ L_{11}^{-2} $.  
Rescaling  \rf{4.1}  by  $L_{11}^2$    to get a dimensionless
  expression   we would  then   get from \rf{4.1} 
($\LL= {L_{11}\ov \llp} \sim L_{11} \LA$)
\be  \la{323}
    \FF=  u_0 {1\ov k} \LL^3  +  u_1 \log \LL  +  u_2    +   \bar \FF(k)\ , \ \ \ \ \ \ \qquad
 \bar \FF =  { {\zeta(3)}\ov \pi^2} k^2  +   
  \sum_{h=2}^{\infty}\   { \dd'_{h}\ov k^{2h-2}} \ .
\ee
Here    $u_0= {2  \ov 3\pi }w$  where $w$ is the coefficient of proportionality in  $ \LA^3 =w \llp^{-3} $
so  that  matching the rational $c'_1$ coefficient  in \rf{322}   requires $w \sim \pi$.\foot{This is indeed the right identification  as follows from
   the discussion in  \ci{Green:1997as,Russo:1997mk} or  the 
   comparison with the coefficient of the corresponding $R^4$ term in 11d  effective action \rf{3.8}. 
   In the present notation $w= \ha \pi $, \ie   $ \LA^3 =\ha \pi  \llp^{-3} $.
}  The terms 
$u_1 \log \LL   + u_2$    come from $s\, \bar {\HH}({s\ov t }) $ term  in \rf{4.1}. 
The coefficients   $\dd'_h$ are related to  $\dd_h$ in \rf{4.2}
by rescaling by some rational factors.

Thus \rf{323}  has the  same     structure  as the sum of  the $\LL^3,\  \log \LL $ and $  - A(k)$   terms  in \rf{322}. 
The missing $\log k$ term  should be  coming from the 1-loop 
 11d supergravity zero mode  normalization contribution   mentioned above  and   is thus not expected to be  captured 
  by this  qualitative  argument     based just  on  the structure of the 4-graviton amplitude. 

Remarkably, $ \bar \FF$ in \rf{323}  has exactly the same form as the leading $\zeta(3) k^2$  term  plus 
the sum of the subleading $1\ov k^{2h-2}$ terms in $- A(k)$  in \rf{2.5},\rf{2.6}.  Furthermore, 
the transcendental  factors of $\pi$  match   between the coefficients $\dd_h\sim \dd'_h $ in \rf{4.2} 
   and $q_h$  \rf{2.6}   in $A(k)$.  
   To  match the  rational coefficient of the $\zeta(3)$ term in $-A(k)$  we would   need an extra factor  of $1\ov 8$, \ie 
   \be 
  {1\ov 8}  L_{11}^2 \AA_4  = {1\ov 8} \FF \ \ \   \to \ \ \ 
    F  \ . \la{355}
  \ee
 The  exact equality  of the rational  factors  in $\dd'_h$  (that should   differ from $\dd_h$ in \rf{4.2} just by  rational  factors) 
     and $q_h$ in \rf{2.6} 
   may be    hard to  expect  a priori   given  the crude  nature   of the
   above  relation between the 1-loop  supergravity amplitude and  the 1-loop  
   partition function on a curved background.  But 
   a  definite 
    mismatch in powers of the Bernoulli number factors   between  \rf{4.2}  and in \rf{2.6}  
   suggests that some other   contributions (in addition to 1-loop 11d supergravity one) 
    may be missing.

   One may wonder whether to match  the  full expression for  the $A(k)$ term in  $F$ in \rf{322} one needs 
   to include   contributions  of   other   M2-brane BPS states  
   to  the M-theory 1-loop partition function.\foot{In particular, one   may consider 
     contributions of M2-branes  wrapped on 2-cycles of $\CP^3$  part of $S^7/\mathbb Z_k$
    (which,  in the   perturbative  10d 
       string  limit,   are   related to the  type IIA string   world-sheet instantons    \ci{Cagnazzo:2009zh,Gautason:2023igo}
       but here  play a role of  massive modes propagating in the loop).
       These   may be the analogs of   M2-branes  wrapped  on   2-cycles  of CY   space  in  \ci{Gopakumar:1998jq}.
 Note  also   that the field strength of the RR  1-form  $A$ in  the $S^7/\ZZ_k$   metric \rf{a10} 
   may be playing the role of the graviphoton strength  in the discussion of \cite{Dedushenko:2014nya}.      }
   By analogy with a  discussion in \cite{Gopakumar:1998ii,Gopakumar:1998jq}  
   one  may conjecture   that this may lead  to a modification 
   of the measure in the  proper-time integral in \rf{C.10} like
   \be
\la{C10}
\sum_{n=1}^{\infty}\int_{0}^{\infty}\frac{d\tau}{\tau^{2}}e^{-\frac{\tau n^{2}}{R_{11}^{2}}}\ \cdots \ \ \   \to \ \ \   
   \sum_{n=1}^{\infty}\int_{0}^{\infty}\frac{d\tau}{\tau} {\mu^2 \ov  \sinh^2 { (\mu^2 \tau)} }\ e^{-\frac{\tau n^{2}}{R_{11}^{2}}}\ \cdots \ , 
\ee
where $\mu$   is a  mass parameter (that may be related to $L_{11}^{-1}$ in the present context,   so  that 
$\mu R_{11}  \sim { 1\ov k}$). 
Then an extra   factor of the Bernoulli numbers  required to match $\dd_h$ in \rf{4.2}   with $q_h$ in \rf{2.6} 
may  come from the expansion 
   \be \la{327}
    \frac{1}{\sinh^{2}\tau} = -\sum_{h=0}^{\infty}\frac{2^{2h}(2h-1)}{(2h)!}\,B_{2h}\,\ta^{2h-2} =\frac{1}{\ta^{2}}-\frac{1}{3}- \sum_{h=2}^{\infty}\frac{2^{2h}(2h-1)}{(2h)!}\,B_{2h}\,\ta^{2h-2} \ .  \ee
  To see  at the heuristic level  how   that may work out  we may start  with  the localization expression for 
  $\bar A(k)$ in \rf{2.5},\rf{2.6} that    has the following integral representation 
    \cite{Hanada:2012si}
\be
\la{4.4}
\bar A(k) \equiv \sum_{h=2}^{\infty}{q_{h}\ov   k^{2h-2}} =
 \int_{0}^{\infty}\frac{dt}{t}\frac{1}{e^{{k  t}}-1}\Big(\frac{1}{\sinh^{2}t}-\frac{1}{t^{2}}+\frac{1}{3}\Big)\ . 
\ee
One may   also rewrite    the full  expression for $A(k)$ in \rf{2.5}   as\foot{
Note    that  using 
$\frac{1}{\sinh^{2}t} = 4\sum_{n=1}^{\infty} n\,e^{-2nt}$ one may also 
 write 
\  $
A(k) = 4\,\sum_{n,m=1}^{\infty}\,n\,\int_{0}^{\infty}\frac{dt}{t}e^{-(m k+2n) t} \, . $
} 
\be\la{329}
A(k) = \int_{0}^{\infty}\frac{dt}{t}\ \frac{1}{e^{kt}-1}\ \frac{1}{\sinh^{2}t} =
 \sum_{n=1}^{\infty}\int_{0}^{\infty}\frac{dt}{t}\,    \frac{1}{\sinh^{2}t} \,  e^{-k \,  n \, t}\, .
\ee
In \rf{329}  we are assuming that  the  evaluation of  the singular terms (corresponding to 
 the last two terms  in the bracket in \rf{4.4}) is done using a suitable regularization.\foot{For instance, 
 with an analytic regularization like  ${dt\ov t}\to {dt\ov t^{1+\eps}}$ one has
 
 $\int_{0}^{\infty}\frac{dt}{t^{1+\eps}}\frac{1}{e^{kt}-1}(\frac{1}{t^{2}}-\frac{1}{3}) = -\frac{1}{6} \big[ {1\ov \eps}  + \log k+\gamma_{\rm E}-\log(2\pi)\big]-\frac{\zeta(3)}{8\pi^{2}}k^{2}+\mc O(\eps)$.
 
The $\log k$ and $k^{2}$ terms here  agree with (\ref{2.5}). The pole term (plus regularization-dependent transcendental constants) 
 should be discarded  as  part of the regularization prescription. 
}
 Here   $k= {L_{11} \ov R_{11}}$  and we may redefine $t\to (L_{11})^{-1 }\tau$ to put  the integral into a similar form as in \rf{C10}.
 
 Eq. \rf{329}    closely resembles the expression in \cite{Gopakumar:1998jq,Dedushenko:2014nya} 
  used 
to  reproduce the coefficients  of the special  protected  $R^{2}{\rm F}^{2h-2}$ terms \ci{Antoniadis:1993ze} 
in the 4d effective action   of    type II   string (compactified on a CY space) 
     from  a conjectured   1-loop  M-theory 
  correction   coming  from    M2-brane  BPS states.  
Indeed, we  may compare the summand in  \rf{329}  with the 4d effective action of a 
 charged scalar of mass $\mm$ (representing an  M2-brane    wrapped on a 2-cycle in CY space ) 
   in a constant self-dual  gauge  field  background 
\be
\la{4.9}
\Gamma (\mm, e {\rm F})  =- \ha \int_{0}^{\infty}\frac{dt}{t}\tr e^{-t\,(\Delta+\mm^{2})} \ \ \sim \ \int_{0}^{\infty}\frac{dt}{t}\frac{1}{\sinh^{2} ( \ha e {\rm F} \, t)  }\, e^{-t \mm^{2}} \ . 
\ee 
Here   $\rm F$ is the  gauge field strength   and the UV divergent term is  assumed to be subtracted out. 
 Specializing to a  BPS state  with  $\mm = e $ and rescaling $t$  one  gets  the integrand as 
$ \frac{1}{\sinh^{2} t'  }\, e^{-   2   \mm  {\rm F}^{-1}\, t'}$.  
Accounting for  multiple  
 wrappings
 corresponds to $\mm \to n \mm $ and summing  over $n$  so that we get 
 \be \la{50}
 \sum_{n=1}^\infty  \Gamma (n \mm, n \mm {\rm F})\  \sim \ \sum_{n=1}^\infty 
 \int_{0}^{\infty}\frac{dt'}{t'}\frac{1}{\sinh^{2} t'   }\,   e^{-   2 n\,   \mm  {\rm F}^{-1}\, t'}\ . 
\ee
 This  matches \rf{329} 
 if  $ 2 \mm  {\rm F}^{-1}$  is identified with  $k$.\foot{This matching 
  is not  totally  unexpected   given that both  functions  were noticed to be 
  related to the topological string amplitudes (cf. \ci{Bershadsky:1993ta,Antoniadis:1993ze}   and  \ci{Drukker:2010nc,Hatsuda:2015oaa}).} 
 Then     the coefficients  in the $1\ov k$ 
 expansion of \rf{329} 
 are  directly related to the coefficients in the expansion of \rf{50}
 in powers of $ {\rm F}$.
 In the present case ${1\ov k}= {R_{11} \ov L_{11}}$  scales as   the square root of the   effective curvature 
of \adsz   and is thus analogous to $\rm F$.\foot{The difference  between \rf{329} or \rf{50}
   and    the attempted  modification    \rf{C10}  is that the sum   over the 11d  KK modes in \rf{C10} 
involves $n^2$ rather than $n$, but  in going from \rf{4.9} to \rf{50}  this is taken care of   by a rescaling of $t$.
For this to be possible  requires the measure in \rf{C10}   to depend on $n$.}

\iffa 
This suggests that while  the 1-loop 11d  supergravity   partition function does capture the structure of the 
large $k$  expansion of the $\LL^0$ term  in \rf{322}, the precise matching of  $A(k)$ term should 
require  computing  the full  M-theory 1-loop partition function that
 should receive contributions from  extended  BPS M2-brane states. 
To  justify  how one  can get the expression  like \rf{50} matching \rf{329}  remains an open problem.   
\fi

\iffa 
\foot{Naively,   the mass of a state that  includes  "point-particle"  KK  11d momentum $n/R_{11}$  contribution  as well as 
the  contribution of  $n'$ times wrapped M2-brane  contribution  would be (cf. \rf{3.1}) 
$M^2(n,n') = { n^2\ov R_{11}^2}  +   c n'  L_{11}^2  {T_2} = { k^2 L_{11}^2 n^2 }  +   c n' {1\ov 4 \pi^2} \LL^3$ 
where  $ R_{11} =  {1\ov k} L_{11} $ and $\LL  = {L_{11} \ov \lp} $   and 
then the 1-loop  contribution   would  read as $\sim \int {dt \ov t^p  } \sum{n,n'}  e^{- M^2(n,n') t } $. 
\fi 
 \iffa 
\foot{
In our context, 
we are after 1-loop M2  brane;  or 1-loop of  particles that are M2 BPS states.
If we have  $\tr \log[  - D^2 + M^2  ]$  it effectively depends on $M/L$  which is dimensionless   where $L$ is curvature scale. 
For a particle   action is $M \int dt$  where $t$ is proper time; 
 for membrane wrapped on  $S^{2}$   $n$ times  we will have 
$M^2 = L^2   n$   (times $4 \pi$ ) .
So total mass will   be 
$M'^2  = L^2 (  4\pi n  +  m^2  k^2  )$ 
where $1/k   =a$  is radius of $S^{1}$. 
So proper-time integral should contain 
$\int \frac{ds}{s}  s^r   \exp [ - M'^2/L^2   s ] ...$
where $r$ is   power depending on dimensions. 
Point is that   $L$ dependence should drop out in 1-loop expression (modulo cutoff dependence). 
G-V redefined   $t$  by power $M'$ to get   linear dependence. 
also,  for them  $M'^2 = Z Z*$,  Z=  central charge. 
}
 \fi 
\iffa 
The idea  is that one factor of Bernoulli is coming from expansion of sinh and another from summation over $m$. 
That should correspond  to sum in \rf{2.5},\rf{2.6}.
We thus need a  right guess to  generalize the 1-loop  graviton amplitude expression in 11d. 
Heuristically, to capture higher genus string contributions  we need 
presumably M2   to wrap higher genus  Riemann surfaces -- not enough to consider just 11d graviton -- beyond  tree and 1-loop order ? 
{\bf  Indeed, what we are to do is to consider M2  on $S^7/Z_k$: \  wrapped on $S^1$ and also  on 2-cycle in $CP^3$.}
Euclidean  string world sheet  instantons on $CP^3$ were discussed  \ci{Cagnazzo:2009zh}
and this is 1/2 BPS state. So this will be analog of maps to CY  in sect 3,4 of  \ci{Gopakumar:1998jq}.
So we just are to replace CY by $CP^3$. 
But it should   somehow be more straightforward than in G-V  ?
Main ingredients are $\chi_h= {(-1)^{h-1} \over  2 h (2 h -2)} B_{h}$   -- Euler number of $M_h$--  moduli space of Riemann surfaces of genus $h$.  And another   factor of $B_{h-1}$  comes  in the process of  counting  constant  maps (see 1st G-V paper). 
So we need to decode and translate G-V  discussion to the present context  to see how we get $A(k)$ as 
1-loop contribution  in M2  theory. 
\fi

\section{
Type IIA string perturbative expansion} 

Let us now compare  the  expansion of  the localization result for free energy 
in the 't Hooft limit  \rf{2.10} with the perturbative expansion  of  the 
effective action  in type IIA  string theory in \adsc  background. 

Let  us start with the  free energy expanded  in large $N$ and large $k$   with   fixed $\l= { N \ov k}$  \rf{2.10} 
and then  expanded  further  in large $\l$  (see \rf{2188},\rf{2.19}). 
Expressing  $F$ in terms of the type IIA string parameters  $\gs$ and $T$ in \rf{a13},\rf{a14} 
we will attempt to  match the result to the perturbative    low-energy or $\alpha'$  expansion of 
  type IIA string effective action  in the  corresponding \adsc background  \rf{a12} order  by order in small $\gs$.  

Using the  original relations  between  the parameters  \rf{a13},\rf{a14}  \ci{Aharony:2008ug} 
\be\la{111}
N = 4\,\sqrt{2\pi}\, T^{5/2}\,\gs^{-1}, \qquad \l = 2T^{2}, \qquad T = \frac{1}{8\pi}{L^{2}\ov \alpha'} \ , 
\ee
where $L$ is the scale in 10d metric  in \rf{a12} we get from \rf{2188}
\be\la{422}
\widetilde F = \frac{1}{384\pi^{2}} {L^{8}\ov \alpha'^{4}\gs^{2}}-\frac{1}{192} \frac{L^{4}}{\alpha'^{2}\,\gs^{2}}
+ \frac{1}{8\pi^2}  \Big[{\zeta(3)\ov \gs^{2}}-2\zeta(2)\Big] \frac{L^{2}}{\alpha'} +\cdots \ . 
\ee
Here the first term  $ {L^{8}\ov \alpha'^{4 }}$ scales as   the  contribution of  the $R$ term 
in the type IIA effective action  while the third  $\frac{L^{2}}{\alpha'} $  term  -- as  the contribution of the $R^4$ term. 
The second $\frac{L^{4}}{\alpha'^{2}}$  contribution  could come   from the 
  ${1\ov \gs^2 \alpha'^{2} }\int d^{10} x \sqrt G \,  R^3$  term in tree-level string effective action   but such term is absent in  type IIA 
10d   string  theory (on supersymmetry grounds).

  This problematic term  is eliminated if one 
takes  into account the shift of $N$ in \rf{32},\rf{A.20}  implying that the relations  between  the gauge theory $(N,\l)$ and 
string theory 
$(\gs, T)$  parameters   take the  modified form \rf{A.21},\rf{A22}.
Note that if we shift $N$   in \rf{32}  as  $N \to N - {k\ov 24} + {a\ov 24} k^{-1}$  with $a\not= 1$ then this term will not be eliminated.\foot{
Let us mention  that 
 instead of expanding in large $N$    one may expand in  the effective  CFT central charge parameter (the coefficient in the 2-point function of the  stress tensor) 
 $c_T\sim \sqrt{k} N + ...$  \ci{Agmon:2017xes,Chester:2018aca,Binder:2019mpb}
 which is naturally  related to the definition of the Newton's constant in the gravitational dual (the  coefficient in the graviton 
 kinetic term).   This  then leads  \ci{Chester:2018aca,Binder:2019mpb}
 to the expected higher derivative terms  avoiding spurious terms like $R^3$
 (we thank S. Chester  for pointing this out).
 Also, an  alternative  shift  of $N$  in relation  to 4d Newton's  constant 
 was used in \ci{Bobev:2022eus}  and shown to lead to simplification of perturbative expansion. 
 In general, the relation  between  string/M-theory parameters and dual  gauge theory ones 
    is effectively a scheme choice required to make the duality manifest; 
    unambiguous  relations are found  only when expressing  one observable in terms of the others. 
 }
As  a result, \rf{2188}  then gives  (cf. \rf{322},\rf{39}) 
\ba
\la{3.10}
&\widetilde F 
= c_{0}\, \frac{L^{8}}{\alpha'^{4}\gs^{2}}  
+\Big(\td c_{0}\,{\zeta(3)\ov \gs^{2}}+c'_{1}\Big)   \frac{L^{2}}{\alpha'}\ - \ \sum_{h=2}^\infty q_{h}\,\Big(\gs^{2}\, \frac{\alpha'}{L^{2}}\,\Big)^{h-1},\\
&c_0= \frac{1}{384\pi^{2}} \ , \ \  \  \td c_0= \frac{1}{8\pi^{2}} \ , \ \ \      c'_1= - {3\ov 64} \ , \ \ \  q_2= -\frac{\pi^{2}}{90}  \ , \ \ \ \ 
q_3= \frac{2\pi^{4}}{2835} \ , \ \ \   q_4= -\frac{8\pi^{6}}{42525} \ , ...\la{310}
\ea
where  we kept only the leading   at  large tension (large $\l$ or small $\alpha'\ov L^2$) 
 contribution   at each order in $\gs^2$ expansion (apart from the $\zeta(3) \gs^{-2}$ term).
  Since according to  \rf{A.19} 
\be {\gs^2\ov 8\pi T} = \gs^{2}\, \frac{\alpha'}{L^{2}} = { 1\ov k^2}  \ , \la{01}\ee
the $q_{h} $ coefficients in \rf{310} are the same as \rf{2.6},\rf{215}  appearing  in 
the large $k$ expansion of $A(k)$ term in \rf{2.7} or in \rf{322}.

\subsection{Transcendentality  structure of  the  coefficients} 

Similarly to the discussion  in  section \ref{sec:local}  above, the first term in (\ref{3.10})  originates  from the   
 supergravity part  ${1\ov \gs^2\alpha'^4 } \int d^{10} x \sqrt {-G}\, (R+\cdots)$ of the tree-level 
10d superstring effective action  evaluated on the \adsc  background (see \rf{B.2}). 

The second  $ {L^2\ov \alpha'}$ term  in \rf{3.10}  has the structure that   corresponds to the  contribution  of the   sum of the 
 tree level 
${1\ov (2\pi)^7 \gs^2\alpha' } {1\ov 8} \zeta(3) \int d^{10} x \sqrt {-G} \, R^4 $ term 
and  1-loop  term ${1\ov  (2\pi)^7 \alpha' } {1\ov 4}  \zeta(2) \int d^{10} x \sqrt {-G} \, ( R^4 +\cdots) $. 
The factors of $\pi$  in  the coefficients match perfectly after 
we account for $\pi^5$   coming from the  volume of \adsc (see \rf{B.3}).
As discussed in more detail in  the next subsection, 
fixing the remaining rational coefficients  requires the information  about the RR   field   strength dependent 
terms in the   corresponding superinvariants  which is not available at the moment. 

The higher order $h \geq 2$ terms in \rf{310}   may originate from local terms in type IIA effective  action of the form 
(note that  $R\sim L^{-2}$ and  the $L^{10}$ factor comes from the 10d volume) 
\be \la{317}
\gs^{2h-2}\, \alpha'^{h-1}  \int d^{10} x \sqrt{-G} \,   \cL_h  \ , \ \ \ \ \ \ 
\cL_h= e_{h,1}  R^{h+4 } + \cdots+ e_{h,r}  D^{2r} R^{h - r + 4 } +\cdots + e_{h,h}D^{2h} R^4 \ . 
\ee
Here  $\cL_h$ may contain  several terms of the same dimension (depending on curvature and   other   fields) 
required on supersymmetry grounds. 
The structure of  these invariants is not  known  but as the relative  coefficients 
of the terms in \rf{317}  should be rational, 
we may get some information about their transcendentality properties    by looking at  the particular 
 terms  $D^{2h} R^4$. 
  Like for   the tree-level  and 1-loop $R^4$ terms,  the 
  coefficients of these terms  may, in principle, be fixed  using 
   the  type II string 4-graviton scattering amplitude.

 In fact,  one  may follow \ci{Russo:1997mk}   and conjecture 
 that in the  perturbative string theory  limit ($\gs \ll 1$) 
     the structure of the 11d supergravity amplitude \rf{4.1} 
   implies the presence of  special 10d-local terms 
   $  \gs^{2h-2} \alpha'^{h-1} \int d^{10} x \sqrt{ -G} \, D^{2h} R^4$    in the  type IIA string effective 
   action. 
   These  should correspond   to local $s^h$ terms in \rf{411}, \ie    should  have the coefficients  
   proportional to $\dd_h$  in \rf{4.2}  or  $ (2\pi)^{2h}$  (after including an overall normalization factor $\sim \pi^2$ as implied by \rf{316}) 
   and  should thus  match the  $\pi$-dependence of the $q_h$ coefficients in \rf{310}. 
   As was  already pointed out    in the previous  section, 
   the matching of  the rational  factors (proportional to $B_{2h} B_{2h-2}$  in \rf{2.6},\rf{215} instead of just $B_{2h-2}$ in $\dd_h$ in \rf{4.2}) 
   implies  the need to account also for the   contributions of other  terms  of the same dimension 
    in the corresponding superinvariants  in \rf{317}.

The same   conclusion  about the structure of the relevant   coefficients 
can be reached also  from the  leading 
$D^{2n} R^4$   terms in  the effective action reconstructed directly from    the type II  4-graviton 10d superstring   amplitudes.
This applies also to the type IIB  effective action 
(for a  related discussion  in connection with  free energy of $\N=2$   4d   gauge theory models  see  \ci{Beccaria:2023kbl}). 
The leading  $D^{2n} R^4$   terms  are the same  (at least at 1-loop and 2-loop orders)   in both type IIA and type IIB  theories \cite{Green:1999pu,Green:2008uj}. 
 In the  type IIB  case 
one finds 
\begin{align}
\la{3.5}
S= {1 \ov (2 \pi)^7} \int d^{10} x \sqrt {-G}   {}&   \Big[\alpha'{}^{-4}  \gs^{-2} R   + \alpha'{}^{-1} \ff_0(\gs)  R^4 
+\sum^\infty_{n=1}\alpha'{}^{n}  \ff_n(\gs)  D^{2n+2} R^4
 \Big] \ . 
\end{align}
The functions $\ff_{0},\ff_1,\ff_2$  contain  a finite number of perturbative contributions plus non-perturbative 
$O(e^{-1/\gs^2})$  corrections  that we shall omit (see, \eg, \ci{Green:1999pu,Green:2005ba,Green:2006gt,
DHoker:2014oxd})\foot{
Here   $\ff_3 = {1\ov 64}  \zeta(9) \gs^{-2}  
  + k_0 \zeta(3) \log( - \alpha' D^2)  + O(g_s^2) $   and may contain   an infinite series of terms  in $g^2_s$ (though their presence 
     appears to remain an open  question). 
The logarithmic term is associated with a  non-local term $p^{16} \log p^2$ 
  in the 4-graviton amplitude on a flat background.}
\begin{align} 
&\ff_0 =\te  {1\ov 8} \Big(  \zeta(3)  \gs^{-2}  + 2 \zeta(2) \Big)  
\ , \qquad \qquad \ \ \ \ff_1 = {1\ov 16}  \Big(   \zeta(5)  \gs^{-2}    +    \tfrac{4  }{3} \zeta(4) \gs^2 \Big) \ , \la{366} \\
&\ff_2 =\te  {1 \ov 48 }  \Big(  [\zeta(3)] ^2  \gs^{-2} + \zeta(3) \zeta(2)  +  6 \zeta(4)    \gs^2  +   \tfrac{2}{9} \zeta(6)    \gs^4\Big)  \ , \ ... \la{3.6}
\end{align}
The leading   $\alpha'$ terms  at each order in $g^2_s$   in \rf{3.5} correspond to the last perturbative terms in $\ff_0,\ff_1,\ff_2$  in \rf{3.6}
and   their coefficients  are  expected to be  protected by supersymmetry. 

This  suggests  that  the coefficients of the  terms 
 $  \gs^{2h-2} \alpha'^{h-1} \int d^{10} x \sqrt{ -G} \, D^{2h} R^4$ with $h\geq 2 $  we are interested in are proportional to  
  $\zeta(2h) =   (-1)^{h+1} ( 2 \pi)^{2h}  {B_{2h} \over 2  ( 2 h)!} \sim ( 2 \pi)^{2h} $.  This
  is the same  conclusion that follows from the above  conjectured relation to the 11d  supergravity 1-loop amplitude.
  Once again, to  match the remaining rational factors  in the coefficients  against those in the free energy \rf{215} would   require the precise knowledge of the  superinvariants   that have the same dimension  as $D^{2h} R^4$  terms.

\subsection{Contributions  from  tree level and 1-loop $R^4$ invariants}

To illustrate this point let us go back to the discussion of the contribution of the  tree-level and 1-loop  $R^4$ terms in type IIA theory.
 They  may be written as (assuming the dilaton is constant and 
  ignoring  dependence on $B_2$ field, see, \eg, \cite{Tseytlin:2000sf} for a review)
 \ba\la{51}
S =& \frac{1}{(2\pi)^{7}}\int d^{10}x\,\sqrt{-G}\, 
\Big[{1\ov  \alpha'^{4} \gs^{2} } (R +\cdots) +  {1\ov \alpha'}   \Big(  {1\ov  \gs^{2}}  r_0  J_0 +  r_1  J_1  \Big) + \cdots \Big]  \ , \\
J_{0} =&\te    t_{8} t_{8}RRRR+\tfrac{1}{8}\eps_{10}\eps_{10}RRRR + \cdots , \qquad 
J_{1} =  t_{8} t_{8}RRRR -\tfrac{1}{8}\eps_{10} \eps_{10}RRRR + \cdots, \la{155} \\
r_{0} =&\te  \frac{1}{3\cdot 2^{11}}\, \zeta(3), \qquad \qquad  
r_{1} =  \frac{\pi^2 }{3\cdot 2^{11}}\,  2  \zeta (2) = \frac{\pi^2 }{3^{2}\cdot 2^{11}} \ .  \la{52}
\ea
Dots in $J_0$ and $J_1$ stand   for other terms of the same dimension  depending on  RR  fields.\foot{Terms  with Ricci tensor can be expressed in terms  of flux-dependent terms using equations of motion (or field redefinitions).
In $J_0, \ J_1$  we use Minkowski signature  so that $\eps_{10}\eps_{10}=-10!$ and after reduction to 8 spatial
dimensions $\eps_{mn\dots}\eps_{mn\dots}\to -2\eps_{8}\eps_{8}$.  $t_{8}$ is the 10-dimensional extension of the 8-dimensional light-cone gauge   involving  $G^{\mu\nu}$ 
(see,  \eg,   \cite{Green:2012pqa}).  Explicitly, \\
$t_{8} t_{8}RRRR = t^{\mu_{1}\nu_{1}\dots\mu_{4}\nu_{4}}t_{\mu_{1}'\nu_{1}'\dots\mu_{4}'\nu_{4}'}R^{\mu_{1}'\nu_{1}'}_{\mu_{1}\nu_{1}}\cdots R^{\mu_{4}'\nu_{4}'}_{\mu_{4}\nu_{4}}, $ \ \   $
\eps_{10}\eps_{10}RRRR = \eps^{\alpha\beta\mu_{1}\nu_{1}\dots\mu_{4}\nu_{4}}\eps_{\alpha\beta\mu_{1}'\nu_{1}'\dots\mu_{4}'\nu_{4}'}R^{\mu_{1}'\nu_{1}'}_{\mu_{1}\nu_{1}}\cdots R^{\mu_{4}'\nu_{4}'}_{\mu_{4}\nu_{4}}$.} 
In type IIB theory $J_1$ is replaced by $J_0$ so that $\frac{\zeta(3)}{\gs^{2}}+\frac{\pi^{2}}{3}$  is the  total coefficient of the $R^4$ terms as
was already indicated in \rf{366}. 
In type IIA  theory $J_0$ and $J_1$ should  correspond to separate  superinvariants.
  
 Explicitly,  the  contribution of the $R^4$ terms   is then 
\ba
\Delta S= \tfrac{1}{(2\pi)^{7}\ 3\cdot 2^{11}\alpha'} \int d^{10}x\,\sqrt{-G}\, 
\Big[ & \te \Big(\frac{\zeta(3)}{\gs^{2}}+\frac{\pi^{2}}{3}\Big)( t_{8}t_{8}RRRR + \cdots) \no \\ 
       &  + \te  \Big(\frac{\zeta(3)}{\gs^{2}}- \frac{\pi^{2}}{3}\Big)  ( \tfrac{1}{8}\eps_{10} \eps_{10}RRRR + \cdots)  \Big]  \ . \la{58}
\ea
To compare  to  the    corresponding $L^2\ov \alpha'$ term in  the free energy \rf{3.10}
we are   to evaluate  \rf{58}  on the \adsc  background.\foot{It is curious  to note that if we did not apply the  redefinition in \rf{A.20}  then the  value of   the coefficient $b_1$  in \rf{3.10},\rf{310} would be 
changed to 
$b_1'= - {1\ov 24}$ so that the coefficient of the $L^2\ov \alpha'$ term in \rf{3.10}   would  become
 ${1\ov 8 \pi^2} (\frac{\zeta(3)}{\gs^{2}}- \frac{\pi^{2}}{3})$ as in \rf{422}
  and is thus  exactly proportional to the  coefficient of the second term in  \rf{58}.} 
As was already mentioned above,   since the background is homogeneous  with   $R\sim F_4^2\sim F_2^2 \sim L^{-2}$ 
and the volume of the \adsc   given  by \rf{B.3}, \ie  
\be 
\int d^{10}x\,\sqrt{-G}\, \Big|_{\rm AdS_4 \times CP^3}  =\te  {4\pi^2 \ov 3} ( {1 \ov 2} L)^4 \cdot   {\pi^3 \ov 6}   L^{6} = {\pi^5 \ov  3^2 \cdot 2^3 } L^{10} \ , \la{59} \ee
the   coefficient of  the $\zeta(3)$ term  in \rf{58}  scales  as 
$\frac{1}{(2\pi)^{7}\cdot 3\cdot 2^{11}} {\pi^5 \ov  3^2 \cdot 2^3}  {L^{2}\ov \alpha'}= { 1 \ov 3^3\cdot 2^{16}}  { 1 \ov \pi^2}  {L^{2}\ov \alpha'}
$.  To match $\td c_0= {1\ov 8 \pi^2}$ in \rf{3.10},\rf{310}  thus  requires an extra rational   factor $3^3\cdot  2^{13}$ 
that should  presumably come  from  the curvature contractions and  other terms in  $J_0,\ J_1$ depending on fluxes. 

A factor of the same order  
  does come  from the 
  Weyl-tensor dependent  part of $J_0$:\foot{Note that $\bar J_0$  vanishes in the case of  undeformed 
  AdS$_5 \times S^5$    background 
  (implying, in particular,  no correction to the radius or free energy, cf. \eg  \ci{Banks:1998nr,Gubser:1998nz,Frolov:2001xr})
   but  does not   vanish on  the \adsc one.}
\ba\la{105}
J_0=  \bar{J}_{0} + \cdots \ , \ \ \ \ \  
 &\bar{J}_{0}     = 3\cdot 2^{8}\,(C^{hmnk}C_{pmnq}C\indices{_{h}^{rsp}}C\indices{^{q}_{rsk}}+\tfrac{1}{2}C^{hkmn}C_{pqmn}C\indices{_{h}^{rsp}}C\indices{^{q}_{rsk}})\ , \\
     &\bar{J}_{0}  \Big|_{\rm AdS_4 \times CP^3}   =   3^4 \cdot 2^{14}  L^{-8} \ . \la{1005}
 \ea
The   difference   from  the required $3^3\cdot  2^{13}$    factor  
 may be  attributed to the contributions 
of  other Ricci tensor dependent terms in $t_8 t_8 RRRR$ and  $\eps_{10} \eps_{10}RRRR$   (discussed  in  Appendix \ref{app:curv}) 
and other  RR field strength dependent terms  in the invariants  $J_0, J_1$   (cf. \cite{Frolov:2001jh}). 

\iffa 
\item ABJM on squashed 3-sphere:
The results of  \cite{Bobev:2022eus}, \cf Eqs.(3.39) and (3.40), show that the dependence on the squashing parameter $b=\sqrt{s}$ in the first two terms
in (\ref{3.13}) -- assuming the old dictionary is ok at least for these two terms -- is 
\be
\la{6.1}
F_{b} = \frac{1}{384\pi^{2}}\frac{1}{\gs^{2}\alpha'^{4}}\Big[1+\frac{(s-1)^{2}}{4s}\Big]
+\frac{1}{8\pi^{2}\alpha'}\frac{\zeta(3)}{\gs^{2}}\Big[1-\frac{(s-1)^{2}(s^{2}+1)}{8s^{2}}\Big]+\cdots,
\ee
The gravitational dual is discussed in  \cite{Martelli:2011fu} finding a suitable solution of $d=4$ $\N=2$ gauged supergravity 
and considering its uplift to a supersymmetric solution of 11d supergravity, see Eq.~(2.4) there. In particular, in the  11d 4-form 
we have the usual AdS 4-volume plus a contribution involving the 4d solution gauge field strength.
In the $d=4$ description, the extra terms in square bracket should be entirely due to the $s$-dependent gauge field, see Eq.~(2.8) in \cite{Martelli:2011fu}.
This is because the the AdS$_{4}$ metric in Eq.~(2.12) of  \cite{Martelli:2011fu} is equivalent to the standard metric
and thus curvature tensors are the same as in $s=1$ case. For instance, the $s$-dependent part in square bracket in the first term of (\ref{6.1}) is due to the 
instanton contribution in Eq.~(2.59) of \cite{Martelli:2011fu}.
What is the 10d geometry obtained after dimensional reduction of the above 11d uplift ?
\end{itemize}
{\small
\begin{verbatim}
 about  matching M-theory  4-point S-matrix from ABJM:
1804.00949, 1808.10554, 1906.07195, 2207.11138
One point is that relating 1-loop amplitude in R-T to 
 -- some  terms may be non-local -- 
like D^4 R^4   is claimed to be absent [ 9808061, 0510027, 1502.0337]
The point probably is that   relevant invariants that contribute 
on ads x S/Z are in any case R^m  and not D^n R^4 that   vanish --
but they together (at given dimension) may be parts of 
one (or several) superinvariants 
 and somehow  their coeffs are captured by zeta(2h-2)  that we see in 1-loop amplitude in R-T....
so  even though D^4 R^4 is not there there will be R^6   with expected zeta(6)   coeff., etc. 
How to put this clearly  will need to be seen. 
But pattern should be similar to the one in  N=2 d=4 case we discussed earlier.
%But  that N^{3/2}   +  q  N^{1/2}    structure is universal as 
%both come from R + RRRR  eff action in 11d 
\end{verbatim}
}
\fi 

\section*{Acknowledgments}
 We thank S. Chester  and S. Giombi  for very useful discussions   and  comments on the draft. 
MB was supported by the INFN grant GSS (Gauge Theories, Strings and Supergravity).
  AAT was supported by the STFC grant ST/T000791/1 
   and also 
  acknowledges A. Sfondrini  for the hospitality at the  Filicudi   workshop on  Integrability in lower-supersymmetry systems in June 2023.


\appendix

\section{Notation and  basic relations}     
\la{apA}

Here we  review  the  relations  between M-theory  and type IIA string theory parameters 
in general and also  in the specific case of the \adsz   background   when they are expressed  in terms of 
 of $N$  and $k$  of  ABJM theory \cite{Aharony:2008ug}.

The action of the   11d supergravity is 
\iffa 
\footnote{Notice that the $2\pi$ power in $\kappa_{11}$ is consistent with \cite{Aharony:2008ug}
and the review \cite{Bagger:2012jb}. As a consequence the M2 tension, first fixed in \cite{Klebanov:1996un} 
and reading $T_{2} =(2\pi)^{2/3}(2\kappa_{11}^{2})^{-1/3}$, takes the
form $T_{2} = \frac{1}{(2\pi)^{2}\ell_{P}^{3}}$. It would be $T_{2} = \frac{1}{(2\pi)\ell_{P}^{3}}$ 
with the alternative  common normalization $2\kappa_{11}^{2}=(2\pi)^{5}\ell_{P}^{9}$ .}
\fi
\ba
\la{A.1}
S_{11} =& \frac{1}{2\kappa_{11}^{2}}\int d^{11}x\,\sqrt{-G}\,\Big(R-\frac{1}{2\cdot 4!}F_{mnk\ell}F^{mnk\ell}+\cdots\Big),  \qquad\qquad 
2\kappa_{11}^{2} = (2\pi)^{8}\, \ell_{P}^{9}\ ,
\ea
where   our  normalization  of 11d Planck length $\llp$ here  is  the same as in, \eg, \cite{Aharony:2008ug,Bagger:2012jb}.
The  M2-brane tension is  then   \cite{Klebanov:1996un} 
\be 
T_{2} = \frac{1}{(2\pi)^{2}\,\ell_{P}^{3}} \ . \la{a2}\ee
Assuming compact $x^{11}$   direction the 11d metric  may be written as 
\be
\la{A.2}
ds^2_{11}  =  e^{-\frac{2}{3}\phi}\,  ds^2_{10} +   e^{\frac{4}{3}\phi}    (dx^{11} +e^{-\phi}A_m dx^m )^{2}, \qquad 
\qquad x^{11}\sim x^{11}+2\pi \RR_{11},
\ee
where, upon reduction to 10d,  $ds^2_{10} $   will be the string frame metric  and   $\phi$  the  dilaton. 
The  constant part of the dilaton is related to string  coupling as $\gs=e^{\phi}$, so that 
 \rf{A.1} reduces to the standard  10d  type IIA supergravity action\foot{It reads  $ \frac{1}{2\kappa_{10}^{2}}\int d^{10}x\,\sqrt{-G}\,  [e^{-2\td \phi} (R+ ...) + ...]$  where $\td \phi$ is non-constant part of the dilaton, 
with constant part   included in $\kappa_{10}$.}
  with 
\be
\la{A.3}
2\pi \RR_{11}\,\frac{1}{\gs^{2}}\,\frac{1}{2\kappa_{11}^{2}} = \frac{1}{2\kappa_{10}^{2}}\ , \qquad \ \ \ \ 2\kappa_{10}^{2} = (2\pi)^{7}\gs^{2}\alpha'^{4} \ . 
\ee
The  M2 brane wrapped on the $x^{11}$ circle   gives the fundamental string  action  with
 the standard tension\foot{In relating M2-brane  action  and  the  fundamental  string action   by this  double dimensional reduction the dilaton factors cancel   \cite{Duff:1987bx}.}  
\be
\la{A.4}
 2\pi\, \RR_{11}\   T_{2} = T_{1}\, , \qquad \qquad T_{1} = \frac{1}{2\pi \alpha'}\ .
 \ee 
 From (\ref{A.3}) and (\ref{A.4}), we then learn  that   in the above notation\foot{These  relations follow from  natural assumption that  11d action does not know  about $\gs$   which enters  only via the dilaton. 
  The resulting   identifications are   also 
 consistent with  relations for  D-brane tensions 
 as   D-brane actions   contain  $e^{-\phi}$ factor, \ie   scale as 
 $1\ov \gs$  for constant dilaton.
 A different  option is to  assume  that relation between 11d and 10d actions involves only non-constant part of the dilaton. Then 
(\ref{A.3}) would not have the $1\ov \gs^{2}$ factor in the l.h.s. and (\ref{A.5}) would then take the form  
(see, \eg,   \cite{Bagger:2012jb})
$ \ell'_{P}=\gs^{1/3}\ell_{s}, \quad   \RR'_{11} = \gs\,\ell_{s}$.}
 \be
 \la{A.5}
\RR_{11}=  \ell_{P} =\ell_{\rm s} \ , \ \ \ \ \  \ell_{\rm s}\equiv \sqrt{\alpha'} \ . 
 \ee
For a constant $\phi$ the effective  radius of the 11-th direction is (as   in   \cite{Witten:1995ex,Townsend:1996xj})
\be  R_{11} = e^{{2\ov 3} \phi} \RR_{11} =  \gs^{2/ 3}  \RR_{11} \ . \la{a7}
\ee
Let us now specialize  to the AdS$_{4} \times S^{7}$  space     supported by the 4-form flux
with $\hat N$ units of  charge (which is the near-horizon limit of the background  sourced by  multiple M2-branes
\ci{Duff:1990xz}) 
\ba
\la{A.7}
 ds^{2}_{11}= L_{11}^{2}\Big(\frac{1}{4}ds^{2}_{\rm AdS_{4}}+ds^{2}_{S^{7}}\Big), \qquad 
 ds^{2}_{\rm AdS_{4}} = dr^{2}+\sinh^{2}r\,d\Omega_{3}^{2}, \qquad 
  F_{4}=dC_{3} \sim \hat N  \,\eps_{4}.
\ea
The flux quantization condition  implies  that 
\be
\la{A.9}
\frac{L_{11}}{\ell_{P}} = \big(2^{5}\pi^{2}\hat N \big)^{1/6}.
\ee
Considering      $\ZZ_k$ quotient  of $S^7$   we get   \cite{Aharony:2008ug}
\ba\la{a10}
&\qquad \qquad \qquad  ds^{2}_{S^{7}/\ZZ_k} =ds^{2}_{\rm CP^{3}} +  {1\ov k^2} (d\vp +  k A)^{2}\, , \qquad \ \ \ \  \vp \equiv  \vp + 2 \pi\\
& ds^2_{{\rm CP}^3}=  {d w^s d \bar w^s\ov 1 + |w|^2}   - { w_r \bar w_s   \ov (1 + |w|^2)^2 } d  w^s d   \bar w^r ,\qquad  \ \  \ 
  dA=   i \Big[{ \delta_{sr} \ov 1 + |w|^2} - {w_s \bar w_r\ov (1 + |w|^2)^2} \Big] d w^r \wedge d \bar w^s ,  
 \no  \ea
  and thus 
  \be \la{a11}
    R_{11} = \gs^{2/3}\, \RR_{11} = \frac{\LLL}{k}\
, \ \ \qquad \ \    \hat N = N k \ , \qquad \ \ \   \frac{\LLL}{\ell_{P}} = \big(2^{5}\pi^{2} N k\big)^{1/6}\ . 
\ee
Here $L_{11}$ and $k$ are the parameters of the 11d  M-theory  background.

Upon dimensional reduction  we then get  the 
 metric  and parameters of 10d  string theory
\ba \la{a12}
&ds_{10}^{2} = L^2 \Big(  {1\ov 4}  ds^{2}_{\rm AdS_{4}}+  ds^{2}_{\rm CP^{3}}\Big),\qquad 
\ \     L_\ads = \ha L \ ,   \\
&\ \ L = \gs^{1/3}\,\LLL\ , \qquad \qquad 
\gs =  \big(\frac{\LLL}{k\, \ell_{P}}\big)^{3/2} \ . \la{a122}
\ea
Expressed   in terms of the  dual   gauge-theory parameters $N$ and $k$  the string coupling and the effective 
dimensionless string tension are 
\ba
\la{a13}
&\gs \equiv \sqrt\pi\, ({2 \ov k} )^{5/4}  {N}^{1/4}  =  \frac{\sqrt\pi\, (2\l)^{5/4}}{N} \ , \ \ \ \ \  \ \ \  \ \ \ \  \l= { N \ov k } 
\ , \\
&T \equiv L_{\ads }^{2}  T_1=    { L^2\ov 8 \pi \alpha'} =\gs^{2/3} \frac{\LLL^{2}}{8\pi \alpha'}
= \frac{\sqrt{\l}}{\sqrt 2}\ , \ \ \la{a14}\\
& \la{A.19}    \frac{\gs^2 }{8\pi\, T }  ={\l^2 \ov N^2} =  {1\ov k^{2}} \ .
\ea
The 
M-theory perturbative expansion  corresponds to large curvature scale or large effective  M2  brane  tension for fixed 
parameter $k$ of the background
   \be \la{a15}
   \LL\equiv {\LLL \ov \llp}\gg 1 \ , \ \ \ \ \ \ \ \ \ \ \   \TT_2 \equiv T_2 \LLL^3\gg 1 \ , \ \ \ \qquad k={\rm fixed} \ , \ee 
\ie  to  the large $N$  limit  with   fixed $k$.
The  10d string perturbative expansion   corresponds to $\gs \ll 1$ for fixed $T$, \ie to the the 't Hooft expansion  in 
the  large $N$,    large $k$ limit   with   fixed $\l= {N \ov k}$. 
 
As was   argued in \cite{Bergman:2009zh},  the presence of the M-theory  correction $R^4 C_3$   implies  the shift 
\be
\la{A.20}
N\to N-\frac{1}{24}\big(k-\frac{1}{k}\big),
\ee
which  modifies  the relation between $L_{11}$ and $N$   in \rf{a11}. 
This leads also to a modification  of the expressions for the 10d string parameters $\gs$ and $T$ in \rf{a13},\rf{a14}
\be
\la{A22}
\gs = \frac{\sqrt\pi\, (2\l)^{5/4}}{N}\,\Big(1-\frac{1}{24\l}+\frac{\l}{24N^{2}}\Big)^{1/4}, \qquad 
T  \equiv {L^2\ov 8 \pi \alpha'} =   \frac{\sqrt{2\l}}{2}\,\Big(1-\frac{1}{24\l}+\frac{\l}{24N^{2}}\Big)^{1/2}\ , 
\ee
or, equivalently,  of how $N$  and $\l$ are expressed in terms of them:
\ba
\la{A.21}
N =& 4\sqrt{2\pi}\, \frac{T^{5/2}}{\gs}\Big(1+\frac{1}{48T^{2}}-\frac{1}{384\pi}\frac{\gs^{2}}{T^{3}}\Big),  \qquad\qquad 
\l = 2T^{2}\,\Big(1+\frac{1}{48T^{2}}-\frac{1}{384\pi}\frac{\gs^{2}}{T^{3}}\Big) \ . 
\ea
%
Note  that the  useful  relation  (\ref{A.19})  remains    unchanged. 


\section{Supergravity   contribution to the  free energy}
\la{app:EH}

To find the leading term in  the  M-theory  effective action  one is to evaluate  the
11d supergravity  action
 \rf{A.1} on the  \adsz background. 
 There is a subtlety here: 
 as this is an ``electric"  solution,   the  sign of the 
 flux $F^2$  term is to be reversed  when evaluating the on-shell  value of the action
 (this is also equivalent to adding a particular boundary term). 
 This then gives  the same   value of the on-shell  action as
 found from the effective 4d action  having  AdS$_4$   as its solution
 (see \ci{Kurlyand:2022vzv,Aguilar-Gutierrez:2022kvk}  and refs. there 
  for related discussions).\foot{This  subtlety is absent in the case of the 
  ``magnetic"  AdS$_7 \times S^4$   solution where  the value 
  of the 11d  action is the same as of the  effective 4d one.}
  The latter approach was used in \ci{Drukker:2010nc}.
  
  Analogous   remark applies to computing  the on-shell  value of the  type IIA  action.
 \iffa 
 More precisely, to get the same  value  as in \ci{Drukker:2010nc}
 one is first to compactify to 4 dimensions, reconstruct 
 effective 4d action that has  AdS$_4$  as its solution and then  compute  the value 
 of this action. Direct  evaluation of 11d action does not give the same  value.
 This is the same phenomenon  as was discussed in \ci{Kurlyand:2022vzv} 
 for the AdS$_5 \times S^5$  case. 
 \fi 
 Explicitly, starting from the  IIA  supergravity solution  and 
compactifying   on $\CP^3$   we get  an  effective 
4d Einstein action with a cosmological constant
 that admits  the AdS$_4$  solution with the radius $L_{\ads}=\ha L$ (cf. \rf{a12}). 
 Then   (using the   negative  overall sign for the action  corresponding to  the Euclidean signature) 
\ba
S_0 = -  \,\frac{1}{2\kappa_{10}^2}  L^6\,\text{vol}({\CP}^{3}) \int d^4 x  \sqrt{g}  \Big( R  + 6 L_\ads^{-2}  \Big)  \ . 
\ea
Here $\kappa_{10}$ is  given in \rf{A.3}  and on the  AdS$_4$  solution $R= - 12L_{\ads}^{-2}$  so that 
\be S_0=- \frac{1}{2\kappa_{10}^{2}}  L^6\,\text{vol}({\CP}^{3})  \,  L_{\ads}^4 \text{vol}(\ads) \, ( - {6}){L_{\ads}^{-2}}
 ={1\ov 3\cdot 2^7 \pi^{2}}\,  {1\ov \gs^{2}}\,  {L^8\ov \alpha'^4 }  = {\pi \sqrt 2 \ov 3}  \l^{-1/2} N^2 \ . 
\label{B.2}
\ee
We used that    the volumes  for the unit-radius  spaces are\foot{In general,  
$\text{vol}({\rm AdS}_{2n}) = \frac{(-2\pi)^{n}}{(2n-1)!!}, \ \ \ 
 \text{vol}({\CP}^{n}) = \frac{1}{2\pi}\text{vol}(S^{2n+1}) =  \frac{\pi^{n}}{n!}$.}
\be \la{B.3}
\text{vol}(\ads) = {4\pi^2 \ov 3} \ , \ \ \ \ \ \qquad    \text{vol}(\CP^3)= {1\ov 2 \pi} \text{vol}(S^3)= {\pi^3 \ov 6} \ . \ee
and also that $L_{\ads}=\ha L$  and \rf{a122},\rf{a14},\rf{a13}. 
Thus \rf{B.2}  matches the first term in $F$ in \rf{2.19}  \ci{Drukker:2010nc}.

The same value   is found by  starting 
with  
the  11d supergravity  solution    and again evaluating the effective 4d action 
 on the 
 \adsz background  
\ba 
S_0 = & - \frac{1}{2\kappa_{11}^{2}}  {1\ov k}  L_{11}^7\,\text{vol}(S^7)  \,  \te ({1\ov 2}L_{11})^4\,  \text{vol}(\ads) \, ( - {6})({1\ov 2}L_{11})^{-2}
\no\\
= & { 1\ov 3 \cdot 2^7 \pi^2 } {1\ov k}  { L_{11}^9\ov \llp^9} = {\pi \sqrt 2 \ov 3}  k^{1/2} N^{3/2} \ , \la{b4}
\ea
where we used \rf{a11}. This  matches the first term in the  large $N$  expansion of free  energy  \rf{2.7}. 
Note that  if  one directly evaluates the  11d action \rf{A.1}  on the  \adsz
solution (without inverting the sign of the $F^2$  term) one gets  $- \ha$ of the value
in \rf{b4}.\foot{Indeed, starting with 11d action  \rf{A.1} 
 with Lagrangian $L= R-  \bar F^2, \ \  \bar F^2 \equiv \frac{1}{2\cdot 4!}F_{mnk\ell}F^{mnk\ell}$  the  corresponding equation  for the metric  
 $R_{mn}  - 4 \bar F_{mklr} \bar F_{n}^{\ \  klr}  - {1\ov 2} G_{mn} (R-  \bar F^2) =0$ 
 implies(using  that  $G^{mn}G_{mn}=11$) that 
  $\bar F^2 = 3 R$  so that the on-shell  value of $L$ is $-2R$. 
  At the same time, reversing the sign of the $\bar F^2$   term gives 
  $R +  \bar F^2 = 4 R$. Here $R=R_4 +R_7= -{12\ov ({1\ov 2}L_{11})^{2}} +  
  {42 \ov (L_{11})^{2}} = - {6 \ov (L_{11})^{2}}$.
  Then  the value of  $      - \frac{1}{2\kappa_{11}^{2}}\int d^{11}x\,\sqrt{-G}
  \,(R+   \bar F^2)$  matches the one in \rf{b4}.}


\section{Values of $R^4$   invariants on $\ads\times \CP^{3}$}
\la{app:curv}

In \rf{1005} we presented the  value of the Weyl-tensor part \rf{105} 
of the $J_0$ invariant in \rf{155} on \adsc   background. 
Here we shall   present  the  values of the  full 
  curvature-dependent parts  of the invariants $J_0$ and $J_1$   keeping also the Ricci tensor  dependent contributions. 

Using the explicit form of the  $t_{8}$  tensor  
 one finds\foot{Compare to  \cite{Schwarz:1982jn}   we do not include $\eps_8$ term in $t_8$, i.e. follow the conventions in  e.g. \ci{Tseytlin:1995bi,Tseytlin:2000sf}.}
\ba\la{c1} 
t_{8}t_{8}RRRR &= t^{mnlrstpq}t_{abcdefgh}R\indices{_{mn}^{ab}}R\indices{_{lr}^{cd}}R\indices{_{st}^{ef}}R\indices{_{pq}^{gh}}\ \\ 
&= -96 R_{ab}{}^{ef} R^{abcd} R_{ce}{}^{gh} R_{dfgh} + 384 
R_{a}{}^{e}{}_{c}{}^{f} R^{abcd} R_{b}{}^{g}{}_{f}{}^{h} R_{dheg} + 
24 R_{ab}{}^{ef} R^{abcd} R_{cd}{}^{gh} R_{efgh}\lp \ \ \ \ 
 - 192 R_{abc}{}^{e} 
R^{abcd} R_{d}{}^{fgh} R_{efgh} 
+ 192 R_{a}{}^{e}{}_{c}{}^{f} R^{abcd} R_{b}{}^{g}{}_{d}{}^{h} 
R_{egfh} + 12 (R_{mnkl} R^{mnkl})^{2}.\no 
\ea
In \rf{155}  one has\foot{In Minkowski signature metric used in \rf{155}  $
 \tfrac{1}{2}\eps_{10} \eps_{10}RRRR  =  - E_8$,  cf. \ci{Tseytlin:2000sf}. } 
\ba 
 & \ \ \ \ \ \ \ \ \  J_0 = t_{8}t_{8}RRRR -  {1\ov 4} E_8 + ... \ , \\ 
E_{8}  \equiv&   
\delta^{n_{1}\cdots n_{8}}_{m_{1}\cdots m_{8}}
R\indices{^{m_{1}m_{2}}_{n_{1}n_{2}}}\cdots R\indices{^{m_{7}m_{8}}_{n_{7}n_{8}}}
\la{c2}  \\  = &-384 R^{2} R_{kl}R^{kl} + 768 (R_{kl}R^{kl})^{2} + 96 
R^{2}R_{klmn}R^{klmn} - 384 R_{pq}R^{pq}R_{klmn}R^{klmn} \lp
+ 
48 (R_{klmn}R^{klmn})^{2}+16 R^{4} 
+R\big(
1024 R_{a}{}^{c} R^{ab} R_{bc} + 1536 R^{ab} R^{cd} R_{acbd} - 1536 
R^{ab} R_{a}{}^{cde} R_{bcde} \lp
- 512 R_{a}{}^{e}{}_{c}{}^{f} R^{abcd} 
R_{bfde}
 + 128 R_{ab}{}^{ef} R^{abcd} R_{cdef}
\big) -1536 R_{a}{}^{c} R^{ab} R_{b}{}^{d} R_{cd} - 6144 R_{a}{}^{c} R^{ab} 
R^{de} R_{bdce}\lp
 + 1536 R^{ab} R^{cd} R_{ac}{}^{ef} R_{bdef}
 + 3072 
R^{ab} R^{cd} R_{a}{}^{e}{}_{c}{}^{f} R_{bedf} 
 + 3072 R_{a}{}^{c} R^{ab} R_{b}{}^{def} R_{cdef} \lp
 - 3072 R^{ab} 
R^{cd} R_{a}{}^{e}{}_{b}{}^{f} R_{cedf} 
+ 6144 R^{ab} R_{a}{}^{cde} 
R_{b}{}^{f}{}_{d}{}^{g} R_{cgef} - 1536 R^{ab} R_{a}{}^{cde} 
R_{bc}{}^{fg} R_{defg}  \lp
+ 3072 R^{ab} R_{a}{}^{c}{}_{b}{}^{d} R_{c}{}^{efg} R_{defg} 
- 1536 R_{a}{}^{e}{}_{c}{}^{f} R^{abcd} R_{b}{}^{g}{}_{e}{}^{h} 
R_{dgfh} - 1536 R_{ab}{}^{ef} R^{abcd} R_{c}{}^{g}{}_{e}{}^{h} 
R_{dhfg} \lp
+ 96 R_{ab}{}^{ef} R^{abcd} R_{cd}{}^{gh} R_{efgh} 
- 768 
R_{abc}{}^{e} R^{abcd} R_{d}{}^{fgh} R_{efgh} + 768 
R_{a}{}^{e}{}_{c}{}^{f} R^{abcd} R_{b}{}^{g}{}_{d}{}^{h} R_{egfh}.\no
\ea
Note that $(R_{klmn}R^{klmn})^{2}$   terms cancel in the combination  of \rf{c1} and \rf{c2} that enters 
$J_0$ (cf. \cite{Myers:1987qx}). 


\def \mum {\gamma}

Evaluating these two invariants  on $\ads\times \CP^{3}$ with   the metric  \rf{a12}  introducing  for generality 
$\mum= ({L_{\CP^3}\ov L_{\rm AdS_4}})^2$\
 as the ratio of the squares of the radii   
\iffa \ba
ds^{2}_{\ads} =& \frac{1}{\mum}\frac{1}{z^{2}}(-dt^{2}+dz^{2}+dx_{1}^{2}+dx_{2}^{2}), \quad
ds^{2}_{\CP^{3}} = \Big[\frac{\sum_{a}dw_{a}d\bar w_{a}}{1+\sum_{a}\bar w_{a} w_{a}}-\frac{\sum_{a,b}w_{a}\bar w_{b}dw_{b}d\bar w_{a}}
{(1+\sum_{a}\bar w_{a} w_{a})^{2}}\Big].
\ea  \fi 
we find\foot{In particular, 
$R_{kl}R^{kl} = (384+36\mum^{2})L^{-4}, \ \  R_{klmn}R^{klmn} = (384+24\mum^{2})L^{-4}, \ \  R^{ab}R_{bc}R^{cd}R_{ca} = (24576+324\mum^{4} )L^{-4}
$.}
\be
t_{8}t_{8}RRRR = 2^{9}\cdot 3^{2}\,(3\mum^{4} +48\mum^{2} + 512) L^{-8} \ , \qquad
E_{8} =  2^{15}\cdot 3^{3}\,\mum\,(3\mum-8) L^{-8} \ . \la{c3}
\ee
For  $\mum =4$ corresponding to  the metric in  \rf{a12} we get 
\be
t_{8}t_{8}RRRR =  2^{20}\cdot 3^{2}   L^{-8} \ , \qquad \ \ \   E_8 =  2^{19}\cdot 3^{3}   L^{-8} \ . 
\ee
Thus   if we would keep only  these curvature-dependent terms  in $J_0, J_1$   in \rf{51}  we would get from \rf{58} using  \rf{59}
\be 
\widebar {\Delta S}= {1 \ov 8 \pi^2}  {L^2\ov   \alpha'} \Big[   \te   {4\ov 3} \Big(\frac{\zeta(3)}{\gs^{2}}+\frac{\pi^{2}}{3}\Big)    
     - {1\ov 2}     \Big(\frac{\zeta(3)}{\gs^{2}}- \frac{\pi^{2}}{3}\Big)     \Big]  \ .
\la{588}
\ee
This is  of the same order  as  just the Weyl-tensor part  contribution  in \rf{1005}   but 
  does not  match  the precise   rational  coefficients in  the 
$ {L^2\ov \alpha'}$ term in $\widetilde F$ in \rf{3.10}.
This suggests that   it is important to include also  the 
contributions of  the RR  field  strength dependent  terms in $J_0$ and $J_1$  to get the matching.

As an aside, let us note that $E_8$ has an interpretation of    an  Euler density  in 8 dimensions.  In general,  for a  $d$ dimensional 
space $\mc M^{d}$ 
with Euclidean signature\foot{Recall that in $d$ dimensions
$
 \eps_{i_{1}\cdots i_{n}}\eps^{j_{1}\cdots j_{n}} = \delta^{j_{1}\cdots j_{n}}_{i_{1}\cdots i_{n}} = \sum_{\sigma}(-1)^{\sigma}\delta^{j_{1}}_{i_{\sigma_{1}}}\cdots\delta^{j_{n}}_{i_{\sigma_{n}}}, \ \ 
 $
$
\delta_{i_{1}\cdots i_{s}i_{s+1}\cdots i_{p}}^{j_{1}\cdots j_{s}j_{s+1}\cdots j_{p}} = \frac{(d-s)!}{(d-p)!}\,\delta_{i_{1}\cdots i_{s}}^{j_{1}\cdots j_{s}} 
$.
}
\be 
E_{2n}(\mc M^{d}) = \frac{1}{(d-2n)!}\,\eps_{d}\eps_{d}R^{n} = \delta^{a_{1}\cdots a_{2n}}_{b_{1}\cdots b_{2n}}R\indices{^{b_{1}b_{2}}_{a_{1}a_{2}}}\cdots R\indices{^{b_{2n-1}b_{2n}}_{a_{2n-1}a_{2n}}},
\ee
which vanishes  for $2n>d$.
For  example, for a  sphere $S^{d}$ we have 
$
R\indices{^{ab}_{ce}} = \frac{1}{r^{2}_d}\delta^{ab}_{ce},
$
and thus $
E_{2n}(S^{d}) = 2^{n} \frac{d!}{(d-2n)!}\big(\frac{1}{r^{2}_d}\big)^{n}.
$
For  a product manifold $\mc M^{m}\times S^{n}$, with $d=m+n$, we have  \cite{Nishida:2013yva}
\be
E_{2p}(\mc M^{m}\times S^{n}) = \sum_{t=0}^{[m/2]}\binom{p}{t}\frac{n!}{(n-2(p-t))!}\Big(\frac{1}{r^{2}_{n}}\Big)^{p-t}\,2^{t-p}E_{2t}(\mc M^{m})
\ . \ee
This is a  special case of the  general relation\foot{We use this  opportunity to point out a misprint in 
eq.(4.1)  in \ci{Tseytlin:2000sf}:    the coefficient of the second term in 

$\qquad E_{8}(\mc M^{4}\times \mc M^{7}) = 4E_{2}(\mc M^{4})\ E_{6}(\mc M^{7})+6 E_{4}(\mc M^{4})\ E_{4}(\mc M^{7})
$ 

 is  6 not 12.  The  value of this coefficient was not, actually  used in    \ci{Tseytlin:2000sf}.}
\ba
E_{2p}(\mc M^{m}\times \mc M^{n}) &=\sum_{t=0}^{[m/2]}\binom{p}{t}E_{2t}(\mc M^{m})E_{2(p-t)}(\mc M^{n})\ , 
\ea
implying, in particular,  that 
$
E_{8}(\mc M^{4}\times \mc M^{6}) = 4E_{2}(\mc M^{4})\ E_{6}(\mc M^{6})+6 E_{4}(\mc M^{4})\ E_{4}(\mc M^{6}).
$
Indeed,  one can check that 
\be
E_{8}(\ads\times \CP^{3}) = 4E_{2}(\ads)\ E_{6}(\CP^{3})+6 E_{4}(\ads)\ E_{4}(\CP^{3}),
\ee
in agreement with  the value of $E_8$ in \rf{c3}.

\section{Non-perturbative corrections  to the ABJM free energy\la{D}} 

Here, for completeness,  we recall  some facts about non-perturbative corrections  to free energy of the ABJM theory.

In  M-theory one   may expect  non-perturbative contributions  to the M2-brane partition function 
  related to 
 membranes wrapping a 3-cycle $\mc C_3$ of 11d space  and thus producing a factor $\sim \exp(- T_2 \text{vol}(\mc C_3))$
 where $T_2$ is M2-brane tension in \rf{a2}.  If  $\mc C_3$  wraps 11d circle then this contribution  may be interpreted as the 
 10d fundamental string instanton  with $T_2 \text{vol}(\mc C_3) \to T_1  \text{vol}(\mc C_2) $  
 where $T_1$ is the string tension  (cf. \rf{A.4}) and  $ \mc C_2$ is a 2-cycle in 10d space on which  the 
   string worldsheet is wrapped. If $\mc C_3$   lies in 10d subspace then the corresponding   contribution is that  of the  D2-brane  instanton. 
   
   In the context of the ABJM theory  one may  thus expect two types of  non-perturbative   contributions to $F$
   proportional to ($n=1,2, ...$) 
   \be\la{d1}
 e^{-2\pi n \sqrt{2\l}} =e^{-2\pi n \sqrt{2}\sqrt{N\ov k}} \ , 
\ee
related to  $\CP^{1}\subset \CP^{3}$  world-sheet instantons \cite{Cagnazzo:2009zh} (cf. \rf{a14}), 
and to ($\ell=1,2, ...$) 
\be\la{d2}
  e^{-   (\pi^6 2^{7} \l^{3})^{1/4}  {\ell  \ov \gs}   } =  e^{-\pi \ell \sqrt{2 }\,   {N\ov \sqrt{\l}} }= e^{-\pi \ell  \sqrt{2}\sqrt{kN}}\  .
\ee
 due to D2-brane instantons on  generalized Lagrangian submanifolds 
 with  topology of ${\rm RP}^{3}\subset\CP^{3}$ \cite{Drukker:2011zy} (cf. \rf{a13}). 
\def \JJ {{\rm J}}

In the Fermi gas approach of \cite{Marino:2011eh} the exact localization expression for the 
ABJM partition function is expressed in terms of the
 grand potential $\JJ(\mu, k)$ of a non-trivial  fermionic system as
\be
\la{73}
Z(N,k) = \frac{1}{2\pi i}\int d\mu \ e^{\, \JJ(\mu,k)-\mu N},
\ee
that may be evaluated at large $N$ by a saddle point method. The grand potential  is given by the sum of the perturbative and non-perturbative parts 
\be
\JJ(\mu,k) = \JJ^{\rm p}(\mu,k)+\JJ^{\rm np}(\mu,k),
\ee
where the perturbative one is 
\be
\JJ^{\rm p}(\mu,k) = \frac{1}{3}C(k)\,\mu^{3}+B(k)\,\mu+A(k), \qquad C(k) = \frac{2}{\pi^{2}k}, \quad B(k) = \frac{k}{24}+\frac{1}{3k}. 
\ee
 Evaluating (\ref{73}) with $\JJ^{\rm p}$  part only 
 gives the  function partition function in   \rf{2.3}    given by the Airy function and the  $e^{A(k)}$   factor.
 
  The non-perturbative part is expected to have the form 
\be
\la{76}
\JJ^{\rm np}(\mu,k) = \sum_{n,\ell}  u_{n,\ell}(k,\mu)\exp\Big[-\Big(\frac{4}{k}n  + 2\ell\Big)\,\mu\Big], 
\ee
where  the two sums  may be interpreted as   accounting  for the 
contributions  of the two types of  instantons  mentioned above.
  Isolating the terms with  $\ell=0$  and  $n=0$  we may write 
\be\la{d7}
\JJ^{\rm np}(\mu,k) = \JJ_{\rm I}(\mu,k) + \JJ_{\rm II}(\mu, k)+\delta \JJ^{\rm np}(\mu, k) \  .   
\ee
Terms with both $\ell>0$ and $n>0$ in \rf{76}  (or "bound state" corrections)  given by  $\delta J^{\rm np}(\mu, k)$ 
were discussed in \cite{Hatsuda:2013gj}. 
Here $\JJ_{\rm I}(\mu,k)$ is  given by 
\be
\JJ_{\rm I}(\mu,k) = \sum_{n=1}^{\infty}d_{n}(k)\,e^{-\frac{4n\mu}{k}}\ ,
\ee
where  $d_{n}$ may be determined  using that the ABJM matrix integral  is dual to the partition function  of 
 topological string theory on ${\rm P}_{1}\times{\rm P}_{1}$. 
 $\JJ_{\rm II}(\mu,k) $ has the following structure for $\mu\gg 1$
\be
\JJ_{\rm II}(\mu,k) = \sum_{\ell=1}^{\infty}\big[a_{\ell}(k)\,\mu^{2}+b_{\ell}(k)\,\mu+c_{\ell}(k)\big]\,e^{-2\ell\mu} \ , 
\ee
where the expansion of the coefficients $a_{\ell}$, $b_{\ell}$, $c_{\ell}$  for  small $k$ follows from the WKB expansion of the Fermi gas
representation  \cite{Marino:2011eh}
$
a_{\ell}(k) = \frac{1}{k}\sum_{m=0}^{\infty}a_{\ell,m}k^{2m},
$  etc. 
Conjectures for the closed form of some of these coefficients were  suggested  in \cite{Hatsuda:2012dt,Calvo:2012du}
and a unifying picture were all $(\ell,n)$ terms in  \rf{76} arise from a refined topological string  representation 
was presented in \cite{Hatsuda:2013oxa}.
The saddle point evaluation of (\ref{73}) sets $\mu\simeq \sqrt{N/C(k)}$. Then, the exponent in (\ref{76}) reproduces the 
expected    weights in \rf{d1} and \rf{d2}   
\be
F^{\rm np} = - \log Z^{\rm np} =  \sum_{n,\ell } {\rm f}_{\ell,n}(N,\l)\,\exp\Big[-\pi\sqrt{2}\Big(2n\sql  + \ell\frac{ N}{\sql}\Big)\Big].
\ee
Recently, the prefactor of the  leading  worldsheet instanton correction to the free energy  \cite{Drukker:2010nc} was
directly computed  on the 10d  string theory side in  \cite{Gautason:2023igo}.


\small 

\bibliography{BT-Biblio}
\bibliographystyle{JHEP-v2.9}
\end{document}